\documentclass[twocolumn,preprintnumbers,amsmath,amssymb,showpacs]{revtex4}
\usepackage{graphicx}
\usepackage{color}
\begin{document}
\title{
$s$-wave Superconductivity due to Suhl-Kondo Mechanism  
in Na$_x$CoO$_2\cdot y$H$_2$O: \\
Effect of Coulomb Interaction and Trigonal Distortion
}

\author{
Keiji {\sc Yada}
}

\address{
Toyota Physical and Chemical Research Institute,
Nagakute-cho, Aichi-gun 480-1192, Japan.
}
\author{
Hiroshi {\sc Kontani}
}
\address{
Department of Physics, Nagoya University,
Furo-cho, Nagoya 464-8602, Japan.
}

\begin{abstract}
To study the electron-phonon mechanism of superconductivity in Na$_x$CoO$_2\cdot y$H$_2$O,
we perform semiquantitative analysis of the electron-phonon interaction (EPI)
between relevant optical phonons (breathing and shear phonons)
and $t_{2g}$ electrons ($a_{1g}$ and $e_g'$ electrons) in the presence of trigonal distortion.
We consider two kinds of contributions to the EPI;
the EPI originating from the Coulomb potential of O ions and that originating from the $d$-$p$ transfer integral
between Co and O in CoO$_6$ octahedron.
We find that the EPI for shear phonons, which induces the interorbital 
hopping of electrons, is large in Na$_x$CoO$_2\cdot y$H$_2$O
because of the trigonal distortion of CoO$_2$ layer.
For this reason,
$T_{\rm c}$ for $s$-wave pairing is prominently enlarged owing to interorbital hopping of Cooper pairs induced by shear phonons,
even if the top of $e_g'$ electron band is close to but below the Fermi level
as suggested experimentally.
This mechanism of superconductivity 
is referred to as the valence-band Suhl-Kondo (SK) mechanism.
Since the SK mechanism is seldom damaged by the Coulomb repulsion,
$s$-wave superconductivity is realized irrespective of large 
Coulomb interaction $U\sim5$ eV at Co sites.
We also study the oxygen isotope effect on $T_{\rm c}$,
and find that it becomes very small due to strong Coulomb interaction.
Finally, we discuss the possible mechanism of anisotropic
$s$-wave superconducting state in Na$_x$CoO$_2\cdot y$H$_2$O, resulting from 
the coexistence of strong EPI and the antiferromagnetic fluctuations.

\end{abstract}


\maketitle

\def\na{Na$_x$CoO$_2$}
\def\a{{\alpha}}
\def\e{{\epsilon}}
\def\i{{ {\rm i} }}
\def\k{{ {\bf k} }}
\def\q{{ {\bf q} }}

\newcommand{\simle}
{\raisebox{-0.75ex}[-1.5ex]{$\;\stackrel{<}{\sim}\;$}}
\newcommand{\simge}
{\raisebox{-0.75ex}[-1.5ex]{$\;\stackrel{>}{\sim}\;$}}

\section{Introduction}
Superconducting layered cobalt oxide Na$_x$CoO$_2\cdot y$H$_2$O ($x\sim0.35$, $y\sim1.3$)
with $T_{\rm c}\sim4.5$ K has attracted considerable attention
since it is the first discovered superconducting cobaltates 
 \cite{takada}.
The parent compound Na$_x$CoO$_2$ reveals rich phase diagram
depending on the Na content $x$ \cite{foo}.
Yokoi {\it et al.} have found that the electronic state of Na$_x$CoO$_2$
drastically changes at the boundary $x\sim0.6$ \cite{yokoi}.
For example, the uniform magnetic susceptibility 
decreases with decreasing temperature for $x\le0.6$,
while it shows a Curie-Weiss-like behavior for $x\ge0.6$ \cite{yokoi,hiroi}.
The quasiparticle spectra in photoemission spectroscopy \cite{pes}
and optical conductivity in infrared spectroscopy \cite{wu}
also decrease with decreasing temperature for $x\le0.6$.
It is noteworthy that
these "weak pseudogap" behavior is also observed in the normal state of
superconducting Na$_x$CoO$_2\cdot y$H$_2$O.
In the superconducting state of Na$_x$CoO$_2\cdot y$H$_2$O,
a sizable decrease of the Knight shift is observed
independently of the direction of magnetic field,
which indicates that the spin singlet superconductivity
is realized \cite{kobayashi,kobayashi_nmr,zheng_nmr}.
A power law behavior in $1/T_1T$ suggests
the large anisotropy in superconducting gap, like a line-node state \cite{fujimoto,ishida,zheng_nqr}.
On the other hand, the decreasing rate of $T_{\rm c}$ due to non-magnetic impurities in this system is much smaller than that of the $d$-wave superconductor,
and it is as small as that of $s$-wave superconductor MgB$_2$ \cite{yokoi_imp}.
This fact suggests that the sign of the superconducting gap function
is unchanged everywhere.

In considering the mechanism and the pairing symmetry of superconductor,
the Fermi surface (FS) topology gives the most important information.
The FSs in Na$_x$CoO$_2$ are composed of $t_{2g}$-orbitals of Co ion,
which split into the $a_{1g}$-orbital and twofold $e_g'$-orbitals
due to crystalline electric field.
The first principle calculation based on local density approximation (LDA)
\cite{singh} had predicted the presence of 
six small hole pockets due to the $e_g'$ bands near the K-points, in addition to
a cylindrical FS due to the $a_{1g}$ band around $\Gamma$-point.
However, such $e_g'$ hole pockets are not observed in ARPES measurements, since the $e_g'$ bands are completely below the Fermi level independently of the Na content  \cite{yang}.
They are also not observed in the bulk-sensitive ARPES
using soft X-ray that has longer escape depth \cite{sato}.
Moreover, the shape of FS does not change by the intercalation of water \cite{shimojima}.
We stress that an estimated specific heat coefficient $\gamma_{\rm est}$
using the $a_{1g}$ Fermi velocity given by ARPES \cite{hasan}
is consistent with experimental vaule $\gamma\sim11$ mJ/mol K$^2$.
If there were $e_g'$ hole pockets, the realized $\gamma$
should be about 3 times greater than the experimental value because of 
the large density of states (DOS) given by the $e_g'$ hole pockets,
as pointed out in Ref. \cite{yada1}.
This fact reinforces the observation by ARPES measurements.

To determine the FS topology theoretically,
present authors have studied the normal electronic state of Na$_x$CoO$_2$
based on the fluctuation-exchange (FLEX) approximation,
which is a self-consistent spin-fluctuation theory \cite{yada1}.
In the FLEX approximation, experimentally observed weak pseudogap behaviors
in the DOS and magnetic susceptibility appear when the top of the $e_g'$-band
is slightly below the Fermi level, that is, the $e_g'$ hole pockets are absent.
We found that the weak pseudogap behaviors originate from
{\bf (i)} antiferromagnetic (AF) fluctuations due to Coulomb interaction, 
and {\bf (ii)} large DOS of the $e_g'$ hole pockets that exist slightly below the Fermi level.
When $e_g'$ hole pockets are present, on the other hand,
ferromagnetic fluctuations are induced by the $e_g'$ hole pockets,
and "anti-pseudogap behavior" appears in the DOS \cite{yada1}.
These results are highly inconsistent with experiments.
Therefore, Na$_x$CoO$_2$ should have a single cylindrical FS around the $\Gamma$-point.

After the discovery of Na$_x$CoO$_2\cdot y$H$_2$O, 
various kinds of superconducting states had been proposed
 \cite{PALee,Baskaran,Ogata-review}.
In particular, possibility of triplet superconducting state
due to Coulomb interaction
were investigated based on the FS with $e_g'$ hole pockets,
by using perturbational theory \cite{nishikawa,yanase} or FLEX approximation \cite{mochizuki,kuroki}.
However, successive experimental efforts had revealed that $e_g'$ hole pockets are absent \cite{shimojima},
and singlet superconducting state is realized \cite{kobayashi,kobayashi_nmr}.
Moreover, ferromagnetic fluctuations are not observed by inelastic 
neutron diffraction in a superconducting single crystal \cite{moyoshi}.
Therefore, we now have to find a mechanism of singlet superconducting 
based on the large $a_{1g}$ FS around the $\Gamma$ point.
If $e_g'$ hole pockets are absent, however,
the expected magnetic fluctuations seem to be too small
to realize unconventional superconductivity \cite{mochizuki}.
Considering the small impurity effect on $T_{\rm c}$ 
 \cite{yokoi_imp}, we should examine the possibility of superconductivity
caused by the electron-phonon interaction (EPI).
In fact, several experimental studies
show the presence of considerable electron-boson coupling in Na$_x$CoO$_2$:
For example, the kink structure was observed in the quasiparticle spectrum in ARPES measurement at $\omega\sim600$ cm$^{-1}$\cite{sato}.
In addition, $\omega$-dependence of scattering rate in infrared spectroscopy shows
a steep upturn at $\omega=500\sim600$ cm$^{-1}$ \cite{wu}.
These experiments suggest the strong electron-boson coupling.
Since the energy of these bosonic modes corresponds to the frequency of relevant optical phonons, EPI is expected to be strong in Na$_x$CoO$_2$.

We have studied the electron-phonon mechanism
for superconductivity in our previous paper \cite{yada2},
by noticing two kinds of the optical modes 
(breathing ($A_{1g}$) and shear ($E_g$) modes)
that strongly couple with $3d$ electrons in Co.
We have found that the shear mode phonon,
which represents the oscillation of O ions parallel to the CoO$_2$ layer,
induces the transition of Cooper pairs between different bands.
Due to this mechanism, a considerably strong pairing interaction for $s$-wave pairing is realized,
which is known as the Suhl-Kondo (SK) mechanism \cite{suhl}.
The SK mechanism works even if $e_g'$ hole pockets are absent,
as long as the top of the $e_g'$-band (valence band) is close to the Fermi level
compared with phonon frequency \cite{yada2}.
In Ref. \cite{yada2},
we have discussed that $s$-wave superconductivity due to 
this ``valence band SK effect'' 
is expected to be realized in water-intercalated Na$_x$CoO$_2\cdot y$H$_2$O
since the top of the valence band is supposed to approach 
the Fermi level \cite{koshibae,ionic}.
We note that extended $s$-wave scenario had been proposed
based on the two concentric $a_{1g}$ FSs model \cite{kukoki-s,mochizuki-s}.

However, we did not take account of the large Coulomb repulsive interaction 
in 3$d$ orbitals of Co in Ref. \cite{yada2},
which works to prevent the $s$-wave pairing.
It is an important problem to elucidate whether or not 
the attractive force due to EPI overcomes the Coulomb repulsion
and an $s$-wave superconductivity can be realized.
For that purpose, we have to know the values of EPI and the Coulomb interaction precisely.
According to a recent first-principle cluster calculation 
based on a quantum chemical ab-initio method, the Coulomb interaction
at Co site in Na$_x$CoO$_2$ is $4\sim5$ eV \cite{landron}.
Also, the mass enhancement factor in M$_x$CoO$_2$ (M=Na, K or Rb)
due to optical phonons at $\sim 0.1$ eV
is estimated to be $m^*/m \sim 2$ by ARPES measurement \cite{sato}. 
However, there is little information about the precise
matrix elements of EPI in Na$_x$CoO$_2$.

In this paper, we quantitatively examine the EPI between $3d$ electrons and relevant optical phonons
that involve the deformation of CoO$_6$ octahedron.
The EPI originates from 
both the change of the Coulomb potential
and that of the transfer integral
due to the displacement of O ions.
We calculate the value of EPI
using the second order perturbation of the transfer integral.
We find that the EPI for shear phonon is considerably increased by the trigonal distortion of CoO$_2$ layer,
which had not been taken into account in our previous study \cite{yada2}.
Thus, $s$-wave superconductivity can be realized in Na$_x$CoO$_2\cdot y$H$_2$O
irrespective of the large Coulomb interaction $U\sim5$ eV.
We also study the oxygen isotope effect on $T_{\rm c}$, and find that 
it becomes very small when $U=4\sim6$ eV.

This paper is organized as follows.
In \S \ref{epi}, we derive the EPI microscopically in case with and without distortion of CoO$_2$ layers.
We discuss the change of EPI due to intercalation of water.
The erroneous result of EPI given in Ref. \cite{yada2} is corrected.
In \S \ref{formulation}, we explain the 3-band model for $t_{2g}$ electron system,
and we derive the linearized gap equation.
In \S \ref{result}, numerical results of the transition temperature 
are obtained by solving the gap equation numerically.
In \S \ref{discussion},
we discuss the robustness of $s$-wave superconductivity over the strong Coulomb interaction.
We also discuss the oxygen isotope effect and the possible 
mechanism of anisotropic $s$-wave superconducting state in Na$_x$CoO$_2\cdot y$H$_2$O.
Finally, results of this paper are summarized in \S \ref{summary}.


\section{electron phonon interaction in $\mbox{Na$_x$CoO$_2\cdot y$H$_2$O}$}\label{epi}
\begin{figure}[htbp]
\begin{center}
\includegraphics[scale=0.5]{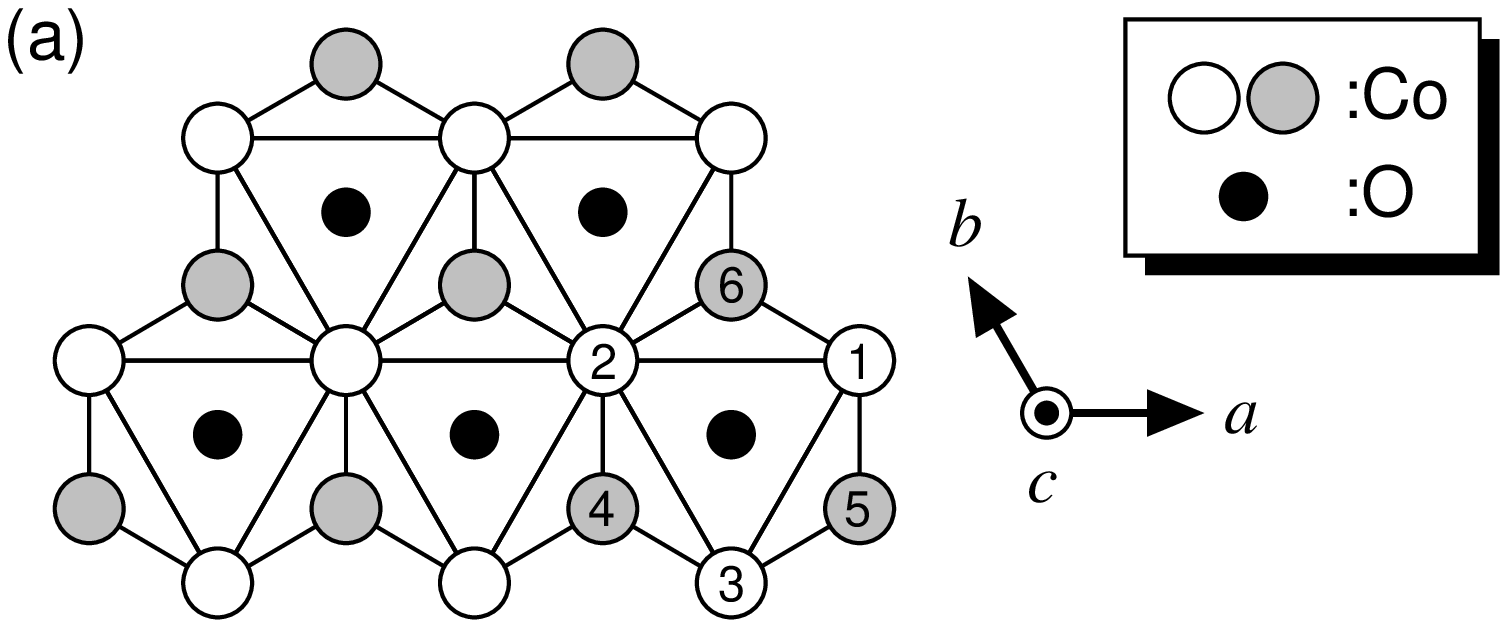}
\includegraphics[scale=0.45]{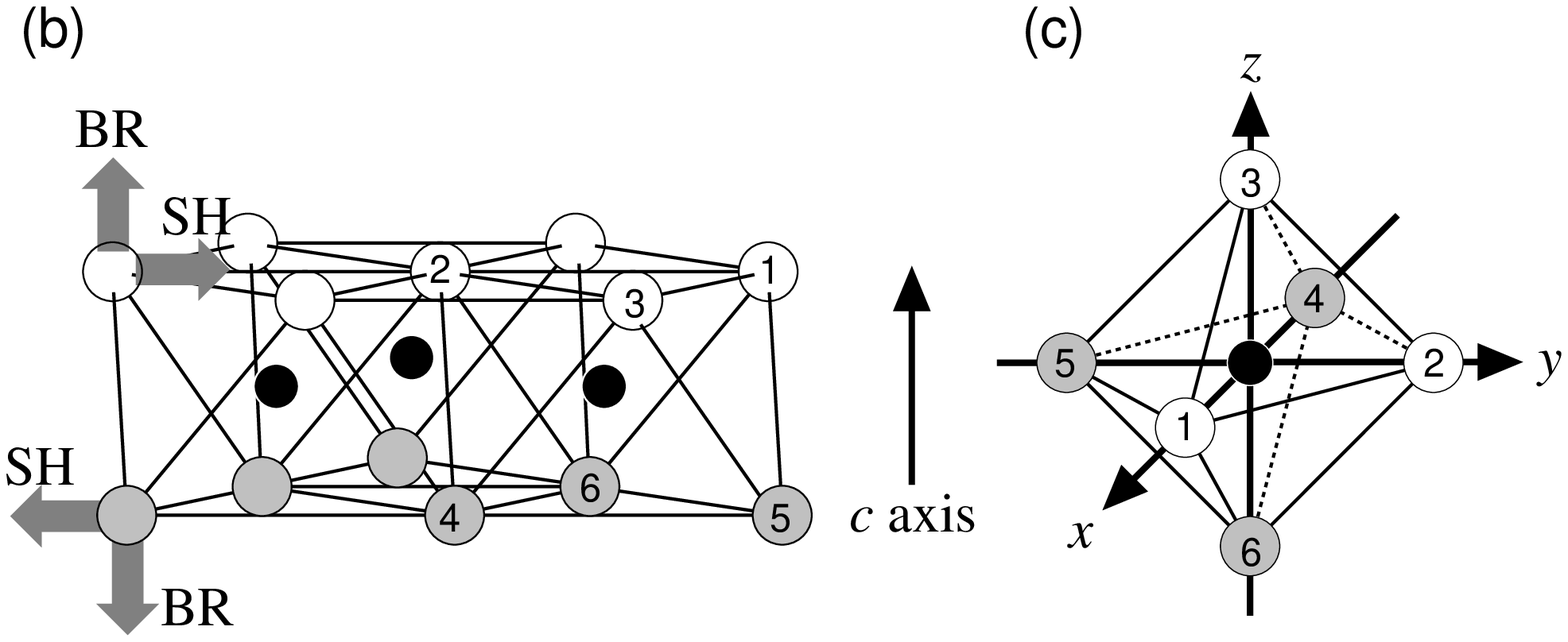}
\caption{(a), (b) Crystal structure of a CoO$_2$ layer.
Co and O in upside (O$_1$, O$_2$, O$_3$) and downside (O$_4$, O$_5$, O$_6$) of CoO$_2$ layer forms triangular lattice, respectively.
(c) The structure of CoO$_6$ octahedron without trigonal deformation.
Note that $z$-axis does not correspond to $c$-axis. }\label{octa}
\end{center}
\end{figure}

\begin{figure}[htbp]
\begin{center}
\includegraphics[scale=0.5]{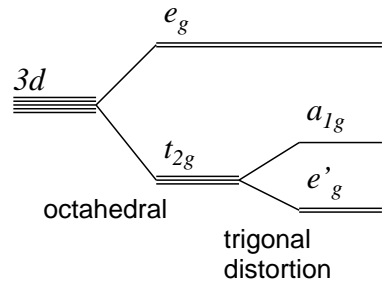}
\caption{Crystalline electronic field for $d$-electron in Na$_x$CoO$_2$.
The splitting between $a_{1g}$ and $e_g'$ is caused by the trigonal deformation of CoO$_6$ octahedron.}\label{cf}
\end{center}
\end{figure}

In this section, we derive the EPI in Na$_x$CoO$_2\cdot y$H$_2$O microscopically.
This compound consists of two dimensional CoO$_2$ layers
separated by a thick insulating layer composed of Na ions and H$_2$O molecules.
CoO$_2$ layer comprises a triangular network of CoO$_6$ octahedra that share edges as shown in Fig. \ref{octa} (a).
In this network, Co ions (black circles in Fig. \ref{octa} (a))
and O ions in the upperlayer (white circles) and lowerlayer (gray circles) forms triangular lattices, respectively.
Due to the crystalline electric field by octahedral coordination of O ion,
fivefold 3$d$ orbitals of Co ion split into threefold $t_{2g}$ ($d_{xy}$, $d_{yz}$ and $d_{zx}$) orbitals
and twofold $e_g$  ($d_{x^2-y^2}$ and $d_{3z^2-r^2}$) orbitals.
Hereafter, we neglect $e_g$-orbitals since they are completely empty.
In Na$_x$CoO$_2$, CoO$_6$ octahedra are trigonally distorted along $c$-axis, so that CoO$_2$ layers become thinner.
Due to this trigonal distortion,
threefold $t_{2g}$ orbitals split into the $a_{1g}$ orbitals and twofold $e_g'$ orbitals
as shown in Fig. \ref{cf}.
According to the LDA calculations, the $a_{1g}$ orbital forms large hole-like FS around the $\Gamma$-point
and the $e_g'$ orbitals form six hole pockets near the K-points.
Since Na ions and O ions are monovalent cation and divalent anion, 
respectively,
valence of Co ions is Co$^{(4-x)+}$ (mixed valence of Co$^{3+}$ and Co$^{4+}$).
Therefore, the number of $3d$ electron is $5+x$ ,that is, the number of hole in $t_{2g}$ orbitals is $1-x$ in Na$_x$CoO$_2\cdot y$H$_2$O.
If the concentration of oxonium ion (H$_3$O$^+$) is $z$, 
the hole number is given by $1-x-z$ \cite{oxonium}.
Since the present paper is devoted to the $s$-wave superconductivity due to the EPI,
obtained results depends not on the hole number,
but on the top of the $e_g'$-bands measured from the Fermi level.
For this reason, we do not consider the existence of the oxonium ion hereafter.

Next, we determine the optical phonon modes that cause strong EPI.
In Na$_x$CoO$_2$, there are 14 kinds of optical modes at zone center \cite{li}.
Since the potentials of $t_{2g}$ orbitals are considerably changed by the deformation of CoO$_6$ octahedron and each CoO$_2$ layer is almost isolated,
we have only to consider single CoO$_2$-layer phonon modes.
According to Ref. \cite{yada2}, there are 4 single-layer phonon modes.
Among them, the ungerade modes in which all O ions move in the same directions
has no couplings with $t_{2g}$ electrons
within the linear term of the displacement.
As a result, only the $A_{1g}$ mode (breathing mode) and the $E_{1g}$ mode (shear mode)
strongly couple with $t_{2g}$ electrons.
In breathing mode and shear mode,
O ions oscillate in the direction parallel and perpendicular to the
CoO$_2$ layer, respectively, as shown in Fig. \ref{octa} (b).

Here, we describe the method of calculation of the EPI.
We focus on the zone center modes and calculate the EPI via the frozen phonon method.
To obtain the EPI, we calculate the change of potentials of $t_{2g}$ orbitals
due to the displacement of six O ions in CoO$_6$ octahedron,
which is shown in Fig. \ref{octa} (c).
We consider two types of contributions to the EPI;
one is the Coulomb potential by O$^{2-}$ ions,
and the other is the effective potential due to the transfer integrals between Co and O.
In Ref. \cite{yada2}, we consider the EPI due to transfer integrals
up to the fourth order processes.
However, the third order term was wrong in sign.
After correcting this mistake,
we verified that a reliable EPI can be obtained by the second order 
perturbation since the third order term and the fourth order term 
almost cancel.
In the present study, we find that the effect of trigonal distortion on EPI,
which was not taken into account in the previous study \cite{yada2},
is significant.
The calculation of the EPI (in the presence of trigonal distortion)
will be shown in \S \ref{epi1} (\S \ref{epi2}) in detail.


\subsection{EPI in the absence of trigonal distortion}\label{epi1}
Here, we calculate the EPI between $t_{2g}$ electrons and optical phonons
in the absence of trigonal distortion of CoO$_6$ octahedron.
We set the $xyz$ coordinate system as shown in Fig. \ref{octa} (c).
$c$ and $a$ crystal axes are along the (1,1,1) and (1,-1,0) directions in the $xyz$ coordinate system, respectively.
The coordinates of six O ions are $(\pm a, 0, 0)$, $(0, \pm a, 0)$, $(0, 0, \pm a)$,
where positive (negative) sign corresponds to O$_1$, O$_2$ and O$_3$ (O$_4$, O$_5$ and O$_6$) ions
that locate upper (lower) side of CoO$_2$ layer.
In breathing and shear modes, O ions oscillate in the direction parallel and perpendicular to the $c$ crystal axis, respectively.
The displacement due to a breathing phonon is ${\bf u}^{\rm BR}=\pm\frac{u^{\rm BR}}{\sqrt3}(1, 1, 1)$,
where positive (negative) sign corresponds to
the displacements of O$_1$, O$_2$ and O$_3$ (O$_4$, O$_5$ and O$_6$) ions.
The displacement due to shear phonon is expressed as 
${\bf u}^{\rm SH}=\pm(v_1, v_2, -v_1-v_2)$.
Here, we choose the orthogonal bases as 
${\bf u}^{\rm SH1}=\pm\frac{u^{\rm SH1}}{\sqrt2}(1, -1, 0)$ and ${\bf u}^{\rm SH2}=\pm\frac{u^{\rm SH2}}{\sqrt6}(1, 1, -2)$,
where positive (negative) sign corresponds to
the displacements of O$_1$, O$_2$ and O$_3$ (O$_4$, O$_5$ and O$_6$) ions.

Here, we calculate the EPI originates from the Coulomb potentials by using the point charge model.
The Coulomb potentials $V_{\rm C}({\bf r})$ of an electron with charge $-e$ 
at the center of six O$^{2-}$ ions is expressed as
$V_{\rm C}({\bf r})=\sum_{i=1,6}\frac{2e^2}{4\pi\varepsilon_0}\frac{1}{|{\bf r-r}_i|}$,
where ${\bf r}_i$ ($i=$1-6) are the position vectors of O ions.
Owing to the displacement of O ions due to $\alpha$-mode phonon,
$V_{\rm C}({\bf r})$ changes to
$V^\alpha_{\rm C}({\bf r})=\sum_{i=1,6}\frac{2e^2}{4\pi\varepsilon_0}\frac{1}{|{\bf r-r}_i-{\bf u}^\alpha_i|}$.
To calculate the EPI, we expand $\delta V_{\rm C}^\alpha({\bf r}) \equiv V^\alpha_{\rm C}({\bf r})-V_{\rm C}({\bf r})$
up to the first order in $u^\alpha$ as follows:
\begin{widetext}
\begin{eqnarray}
\delta V_{\rm C}^{\rm BR}({\bf r})&=&\frac{2e^2}{4\pi\varepsilon_0}
\left(
-\frac{2\sqrt3}{a^2}+\frac{4\sqrt3}{a^4}(xy+yz+zx)
\right)u^{\rm BR},\label{eq:vcbr}\\
\delta V_{\rm C}^{\rm SH1}({\bf r})&=&\frac{2e^2}{4\pi\varepsilon_0}
\left(
-\frac{9\sqrt2}{2a^4}(x^2-y^2)-\frac{3\sqrt2}{a^4}(yz-zx)
\right)u^{\rm SH1},\label{eq:vcsh1}\\
\delta V_{\rm C}^{\rm SH2}({\bf r})&=&\frac{2e^2}{4\pi\varepsilon_0}
\left(
-\frac{3\sqrt6}{2a^4}(x^2+y^2-2z^2)+\frac{\sqrt6}{a^4}(2xy-yz-zx)
\right)u^{\rm SH2},\label{eq:vcsh2}
\end{eqnarray}
\end{widetext}
where we have dropped the fourth and higher order terms in $x$, $y$ and $z$.
The matrix elements of the EPI due to Coulomb potential is given by
$\langle\mu|\delta V_{\rm C}^\alpha|\nu\rangle$, where 
$\mu,\nu=x,y, z$.
Since the wave function for $\mu\nu$-orbital ($\mu\nu=xy,yz,\ \mbox{or} \ zx$) 
is expressed as $\phi_{\mu\nu}({\bf r})=\sqrt{15}\frac{\mu\nu}{r^2}\phi(r)$,
it is easy to show that 
$\langle yz|xy|zx \rangle=\frac17 \langle r^2\rangle$,
where $\langle r^2\rangle=\int \phi^2(r) d^3r$ is the expectation value 
of the square of the radius in $3d$-orbital.
We use the notation $r_d^2=\langle r^2\rangle$ hereafter.
Similarly, $\langle zx|yz|xy\rangle=\langle xy|zx|yz\rangle=\frac{1}{7}r_d^2$.
Thus, we obtain the EPI for breathing phonon as
\begin{widetext}
\begin{eqnarray}
\hat {\delta V}_{\rm C}^{\rm BR}=\frac{2e^2}{4\pi\varepsilon_0}\left\{
-\frac{2\sqrt3}{a^2}\left[
\begin{array}{ccc}
1&0&0\\
0&1&0\\
0&0&1
\end{array}
\right]
+\frac{4\sqrt3r_d^2}{a^4}\left[
\begin{array}{ccc}
0&\frac{1}{7}&\frac{1}{7}\\
\frac{1}{7}&0&\frac{1}{7}\\
\frac{1}{7}&\frac{1}{7}&0
\end{array}
\right]
\right\}u^{\rm BR},
 \nonumber \label{eq:cbr}
\end{eqnarray}
\end{widetext}
where the first, the second and the third column (row) correspond to $d_{xy}$, $d_{yz}$ and $d_{zx}$ orbitals, respectively.

Here, we change the basis of $t_{2g}$-orbitals
from the ($d_{xy}$, $d_{yz}$, $d_{zx}$)-basis into the ($a_{1g}$, $e_g'^1$, $e_g'^2$)-basis,
which is the basis of the irreducible representation under the trigonal crystalline electric field.
The transformation matrix $\hat U$ from the $(d_{xy}, d_{yz}, d_{zx})$-basis to the $(a_{1g}, e_g'^1, e_g'^2)$-basis is given by
\begin{eqnarray}
\left(
\begin{array}{c}
|a_{1g}\rangle\\
|e_g'^1\rangle\\
|e_g'^2\rangle
\end{array}
\right)
&=&
\hat U
\left(
\begin{array}{c}
|xy\rangle\\
|yz\rangle\\
|zx\rangle
\end{array}
\right)
 \nonumber \\
&=&
\left(
\begin{array}{ccc}
\frac{1}{\sqrt3}&\frac{1}{\sqrt3}&\frac{1}{\sqrt3}\\
0&\frac{1}{\sqrt2}&-\frac{1}{\sqrt2}\\
\frac{2}{\sqrt6}&-\frac{1}{\sqrt6}&-\frac{1}{\sqrt6}
\end{array}
\right)
\left(
\begin{array}{c}
|xy\rangle\\
|yz\rangle\\
|zx\rangle
\end{array}
\right).
\end{eqnarray}
By operating $\hat U$ and $\hat U^\dag$ on the left side and the right side of the matrix in Eq. (\ref{eq:cbr}), respectively,
the matrix form of the EPI in the ($a_{1g}, e_g'^1, e_g'^2$)-basis is given by
\begin{eqnarray}
\hat {\delta V}_{\rm C}^{\rm BR}=\left(
\begin{array}{ccc}
a_1^{\rm C}&0&0\\
0&a_2^{\rm C}&0\\
0&0&a_2^{\rm C}
\end{array}
\right)\tilde u^{\rm BR},
\end{eqnarray}
\begin{eqnarray}
a_1^{\rm C}&=&-\frac{2e^2}{4\pi\varepsilon_0}\cdot\frac{2\sqrt3}{a^2}\sqrt\frac{\hbar}{2M\omega_{\rm BR}}\left(1-\frac{4r_d^2}{7a^2}\right),\\
a_2^{\rm C}&=&-\frac{2e^2}{4\pi\varepsilon_0}\cdot\frac{2\sqrt3}{a^2}\sqrt\frac{\hbar}{2M\omega_{\rm BR}}\left(1+\frac{2r_d^2}{7a^2}\right).
\end{eqnarray}
Here, we introduce a dimensionless parameter $\tilde u^\alpha=u^\alpha/\sqrt{\hbar/2M\omega_\alpha}$
that represents a displacement due to $\alpha$-mode phonon as the unit of zero-point motion $\sqrt{\hbar/2M\omega_\alpha}$,
where $\omega_\alpha$ is the frequency of $\alpha$-mode phonon.
Then, $\tilde u^\alpha$ is expressed by the sum of creation and annihilation operators for the $\alpha$-mode phonon,
$\tilde u^\alpha=b_{\alpha}+b^\dag_{\alpha}$.

Next, we calculate the EPI for shear phonons.
Using $\langle xy|x^2|xy\rangle=\langle zx|x^2|zx\rangle=\frac{3}{7}r_d^2$ and $\langle yz|x^2|yz\rangle=\frac{1}{7}r_d^2$,
we obtain the matrix form of the EPI for shear phonons 
from Eqs. (\ref{eq:vcsh1})-(\ref{eq:vcsh2}) 
in the $(d_{xy}, d_{yz}, d_{zx})$-basis as
\begin{widetext}
\begin{eqnarray}
\hat {\delta V}_{\rm C}^{\rm SH1}&=&\frac{2e^2}{4\pi\varepsilon_0}\left\{
-\frac{9\sqrt2r_d^2}{2a^4}\left[
\begin{array}{ccc}
0&0&0\\
0&-\frac{2}{7}&0\\
0&0&\frac{2}{7}
\end{array}
\right]
-\frac{3\sqrt2r_d^2}{a^4}\left[
\begin{array}{ccc}
0&-\frac{1}{7}&\frac{1}{7}\\
-\frac{1}{7}&0&0\\
\frac{1}{7}&0&0
\end{array}
\right]
\right\}u^{\rm SH1}\nonumber\\
&=&
\frac{2e^2}{4\pi\varepsilon_0}\cdot
\frac{3\sqrt2r_d^2}{7a^4}\left[
\begin{array}{ccc}
0&1&-1\\
1&3&0\\
-1&0&-3
\end{array}
\right]
u^{\rm SH1},\label{eq:vc2sh1}\\
\hat {\delta V}_{\rm C}^{\rm SH2}&=&\frac{2e^2}{4\pi\varepsilon_0}\left\{
-\frac{3\sqrt6r_d^2}{2a^4}\left[
\begin{array}{ccc}
\frac{4}{7}&0&0\\
0&-\frac{2}{7}&0\\
0&0&-\frac{2}{7}
\end{array}
\right]
+\frac{\sqrt6r_d^2}{a^4}\left[
\begin{array}{ccc}
0&-\frac{1}{7}&-\frac{1}{7}\\
-\frac{1}{7}&0&\frac{2}{7}\\
-\frac{1}{7}&\frac{2}{7}&0
\end{array}
\right]
\right\}u^{\rm SH2}\nonumber\\
&=&
\frac{2e^2}{4\pi\varepsilon_0}\cdot
\frac{\sqrt6r_d^2}{7a^4}\left[
\begin{array}{ccc}
-6&-1&-1\\
-1&3&2\\
-1&2&3
\end{array}
\right]
u^{\rm SH2},\label{eq:vc2sh2}
\end{eqnarray}
\end{widetext}
By transforming Eqs. (\ref{eq:vc2sh1})-(\ref{eq:vc2sh2}) into the expression in the $(a_{1g}, e_g'^1, e_g'^2)$-basis,
we obtain
\begin{eqnarray}
&&\hat {\delta V}_{\rm C}^{\rm SH1}=\left(
\begin{array}{ccc}
0&b_1^{\rm C}&0\\
b_1^{\rm C}&0&-b_2^{\rm C}\\
0&-b_2^{\rm C}&0
\end{array}
\right)\tilde u^{\rm SH1},
\\
&&\hat {\delta V}_{\rm C}^{\rm SH2}=\left(
\begin{array}{ccc}
0&0&-b_1^{\rm C}\\
0&b_2^{\rm C}&0\\
-b_1^{\rm C}&0&-b_2^{\rm C}
\end{array}
\right)\tilde u^{\rm SH2},
\end{eqnarray}
\begin{eqnarray}
b_1^{\rm C}&=&\langle a_{1g}|\delta V_{\rm C}^{\rm SH1}|e_g'^1\rangle
\nonumber \\
&=&\frac{2e^2}{4\pi\varepsilon_0}\cdot\frac{2\sqrt3}{a^2}\sqrt\frac{\hbar}{2M\omega_{\rm SH}}\frac{4r_d^2}{7a^2},
\\
b_2^{\rm C}&=&\langle e_g'^1|\delta V_{\rm C}^{\rm SH2}|e_g'^1\rangle=b_1^{\rm C}/(4\sqrt2).
\end{eqnarray}
Thus, the EPI originating from the Coulomb potential is represented by four coupling constants $a_1^{\rm C}$, $a_2^{\rm C}$, $b_1^{\rm C}$ and $b_2^{\rm C}$.

\begin{figure}[htbp]
\begin{center}
\includegraphics[scale=0.5]{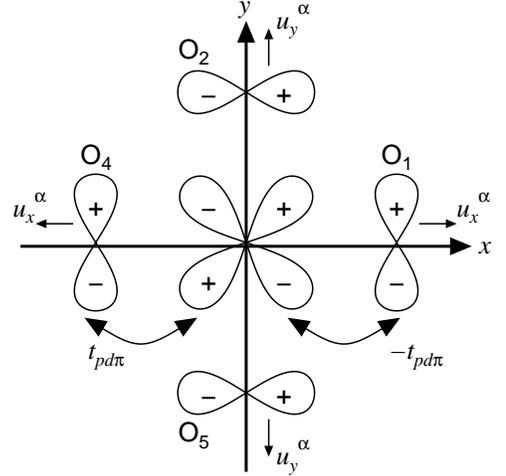}
\caption{Transfer integrals between $d_{xy}$ orbital in Co and $(p_x,p_y)$ orbitals in O.
In this figure, $d_{xy}$ electron can transfer to $p_y$ orbitals
in O$_1$ and O$_4$ via $\pm t_{pd\pi}$.}\label{tpdp}
\end{center}
\end{figure}

Next, we derive the EPI that originates from the changes of $d$-$p$ transfer integrals between Co and O.
For this purpose, we consider the CoO$_6$ octahedron
in which Co$^{4+}$ ion (hole number is unity) is surrounded by six O$^{2-}$ (no holes) ions.
In this octahedron, the uncertainly principle allows the virtual process
in which a hole in $d$ orbitals transfers to $p$ orbitals and turns back.
By this second-order process, the effective potential of a hole in $d$ orbitals is lowered by $-t_{pd}^2/\Delta_{pd}$,
where $t_{pd}$ is the $d$-$p$ transfer integral and $\Delta_{pd}(>0)$ is the charge transfer energy.
When the transfer integral is changed by $\delta t_{pd}$,
the effective potential of a hole becomes 
$-(t_{pd}+\delta t_{pd})^2/\Delta_{pd}$.
Thus, in the electron representation,
the change of the effective potential is given by
$\delta V_{\rm T}=2t_{pd}\delta t_{pd}/\Delta_{pd}$.
In CoO$_6$ octahedron,
the total change of the effective potential
$(\delta V_{\rm T})_{ll'}$ for $d$-orbital is given by
\begin{widetext}
\begin{eqnarray}
(\delta V_{\rm T}^\alpha)_{ll'}=\sum_{i=1-6}\sum_{m=p_x,p_y,p_z} \frac{t_{pd}(i,l,m)\delta t_{pd}^\alpha(i,l',m)+t_{pd}(i,l',m)\delta t_{pd}^\alpha(i,l,m)}{\Delta_{pd}},\label{eq:vt}
\end{eqnarray}
\end{widetext}
where $t_{pd}(i,l,m)$ represents the $d$-$p$ transfer integrals between 
$l$ orbital of Co $(l=d_{xy}, d_{yz}, d_{zx})$
and $m$ orbital of O $(m=p_x, p_y, p_z)$ at $i$ site ($i=1\sim6$),
and $\delta t_{pd}^\alpha(i,l,m)$ is the change of $t_{pd}(i,l,m)$ due to $\alpha$-mode phonon.
In the absence of trigonal distortion,
$t_{pd}(i,l,m)$ is given by Slater-Koster parameter $t_{pd\pi}$ only, as shown in Fig. \ref{tpdp}.
For example, $t_{pd}(i=1,l=d_{xy},m=p_y)=-t_{pd\pi}$ and $t_{pd}(i=4,l=d_{xy},m=p_y)=t_{pd\pi}$.
According to Harrison's law \cite{Harrison}, $t_{pd\pi}$ is proportional to $a^{-4}$,
where $a$ is the distance between Co and O.
Then, $\delta t_{pd\pi}=-t_{pd\pi}\frac{4}{a}\delta a$.

Now, we calculate $(\delta V_{\rm T}^{\rm BR})_{ll}$.
Since $\delta a=u^{\rm BR}/\sqrt 3$ for breathing phonon,
$\delta t_{pd}^{\rm BR}(i,l,m)=-t_{pd}(i,l,m)\frac{4}{\sqrt3a}u^{\rm BR}$.
For each $l$, $t_{pd}(i,l,m)$ is finite only for four sets of $(i,m)$.
After taking the summation of $i$ and $m$ in Eq. (\ref{eq:vt}),
we obtain $(\delta V_{\rm T}^\alpha)_{ll}=-8t_{pd\pi}^2u^{\rm BR}/(\sqrt3\Delta_{pd}a)\times4=-32t_{pd\pi}^2u^{\rm BR}/(\sqrt3\Delta_{pd}a)$.
By performing similar analysis, we obtain
$(\delta V_{\rm T}^{\rm BR})_{ll'}=4t_{pd\pi}^2u^{\rm BR}/(\sqrt3\Delta_{pd}a)$ for $l\neq l'$ by using the Slater Koster formula \cite{slater},
without necessity to use Harrison's law.
As a result, the matrix representation for breathing phonon 
in the $(d_{xy}, d_{yz}, d_{zx})$-basis is expressed as follows:
\begin{eqnarray}
\hat {\delta V}_{\rm T}^{\rm BR}=\left[
\begin{array}{ccc}
 8 & -1 & -1 \\
 -1 & 8 & -1 \\
 -1 & -1 & 8 
\end{array}
\right]\frac{4t_{pd\pi}^2}{\sqrt3\Delta_{pd}a}u^{\rm BR}.\label{eq:vtbr}
\end{eqnarray}
Equation (\ref{eq:vtbr}) is transformed in the 
$(a_{1g}, e_g'^1, e_g'^2)$-basis as
\begin{eqnarray}
\hat {\delta V}_{\rm T}^{\rm BR}=\left(
\begin{array}{ccc}
a_1^{\rm T}&0&0\\
0&a_2^{\rm T}&0\\
0&0&a_2^{\rm T}
\end{array}
\right)\tilde u^{\rm BR},\\
a_1^{\rm T}=\frac{2}{3}a_2^{\rm T}=-\frac{24}{\sqrt3}\cdot\frac{t_{pd\pi}^2}{\Delta_{pd}a}\sqrt\frac{\hbar}{2M\omega_{\rm BR}}.
\end{eqnarray}

In the same way, we can derive the EPI for shear phonons.
For SH1 mode (${\bf u}_{\rm SH1}=\pm\frac{u^{\rm SH1}}{\sqrt2}(1, -1, 0)$), $\delta a=u/\sqrt2$ for O$_1$ and O$_4$,
$\delta a=-u/\sqrt2$ for O$_2$ and O$_5$,
and $\delta a=0$ for O$_3$ and O$_6$.
Therefore,
$\delta t_{pd}^{\rm SH1}(i,l,m)=-t_{pd}(i,l,m)\frac{4}{\sqrt2a}u^{\rm SH1}$ for $i=1, 4$,
$t_{pd}(i,l,m)\frac{4}{\sqrt2a}u^{\rm SH1}$ for $i=2, 5$,
and 0 for $i=3, 6$.
For SH2 mode (${\bf u}_{\rm SH2}=\pm\frac{u^{\rm SH2}}{\sqrt6}(1, 1, -2)$),
$\delta t_{pd}^{\rm SH2}(i,l,m)=-t_{pd}(i,l,m)\frac{4}{\sqrt6a}u^{\rm SH1}$ for $i=1, 2, 4, 5$,
and $t_{pd}(i,l,m)\frac{8}{\sqrt6a}u^{\rm SH1}$ for $i=3, 6$.
As a result, we obtain
\begin{eqnarray}
\hat {\delta V}_{\rm T}^{\rm SH1}=\left[
\begin{array}{ccc}
0&1&-1\\
1&8&0\\
-1&0&-8
\end{array}
\right]\frac{\sqrt2t_{pd\pi}^2u^{\rm SH}}{\Delta_{pd}a}\label{eq:vtsh1}\\
\hat {\delta V}_{\rm T}^{\rm SH2}=\left[
\begin{array}{ccc}
-16&-1&-1\\
-1&8&2\\
-1&2&8
\end{array}
\right]\sqrt{\frac{2}{3}}\frac{t_{pd\pi}^2u^{\rm SH}}{\Delta_{pd}a}\label{eq:vtsh2}
\end{eqnarray}
By transforming Eqs. (\ref{eq:vtsh1})-(\ref{eq:vtsh2}) into $(a_{1g}, e_g'^1, e_g'^2)$-basis, we obtain
\begin{eqnarray}
&&\hat {\delta V}_{\rm T}^{\rm SH1}=\left(
\begin{array}{ccc}
0&b_1^{\rm T}&0\\
b_1^{\rm T}&0&-b_2^{\rm T}\\
0&-b_2^{\rm C}&0
\end{array}
\right)\tilde u^{\rm SH1},
 \\
&&\hat {\delta V}_{\rm T}^{\rm SH2}=\left(
\begin{array}{ccc}
0&0&-b_1^{\rm T}\\
0&b_2^{\rm T}&0\\
-b_1^{\rm T}&0&-b_2^{\rm T}
\end{array}
\right)\tilde u^{\rm SH2},\\
& &b_1^{\rm T}=\frac{3}{\sqrt2}b_2^{\rm T}=6\sqrt3\cdot\frac{t_{pd\pi}^2}{\Delta_{pd}a}\sqrt\frac{\hbar}{2M\omega_{\rm SH}}.
\end{eqnarray}
Thus, the EPI originating from $d$-$p$ transfer integrals is represented by four coupling constants $a_1^{\rm T}$, $a_2^{\rm T}$, $b_1^{\rm T}$ and $b_2^{\rm T}$.

\subsection{EPI in the presence of trigonal distortion}\label{epi2}
In \S \ref{epi1}, we have calculated the EPI in the absence of trigonal distortion.
However, CoO$_2$ layer in Na$_x$CoO$_2$ becomes thinner due to the trigonal 
distortion, and it is increased by the water intercalation.
Since this change of crystal structure can modify the EPI prominently,
we calculate the EPI in the presence of trigonal distortion in this subsection.

Owing to the trigonal distortion, the position of O ions move along the $c$ crystal axis.
Then, the changes of the coordinates of O$_1$, O$_2$ and O$_3$ are $(-b/\sqrt3, -b/\sqrt3, -b/\sqrt3)$
and that for O$_4$, O$_5$ and O$_6$ are $(b/\sqrt3, b/\sqrt3, b/\sqrt3)$,
where $b\ (>0)$ is the magnitude of displacement.
According to the neutron scattering measurement \cite{lynn},
O$_1$-Co-O$_5$ angle $\theta_{15}$ for unhydrated Na$_x$CoO$_2$ and hydrated Na$_x$CoO$_2\cdot y$H$_2$O
are 84$^\circ$ and 82$^\circ$, respectively.
Thus, we can determine the value of $b$ from $\theta_{15}$ by solving the following equation.
\begin{eqnarray}
\cos\theta_{15}=-(3m^2-2m)/(3m^2-2m+1),\label{eq:b}
\end{eqnarray}
where $m=b/(\sqrt3a)$.
Using the values of $b$ obtained in Eq. (\ref{eq:b}), we derive the EPI in the presence of trigonal distortion as follows.

First, we calculate the EPI originating from the change of the Coulomb potential.
In the case of $m=0$, the $u^\alpha$-linear terms of
the Coulomb potentials are given in Eqs. (\ref{eq:vcbr})-(\ref{eq:vcsh2}).
Up to the first order of $m=b/(\sqrt3a)$, they are modified as 
\begin{widetext}
\begin{eqnarray}
\delta V_{\rm C}^{\rm BR}({\bf r})&=&\frac{2e^2}{4\pi\varepsilon_0}
\left(
-\frac{2\sqrt3}{a^2}+\frac{4\sqrt3}{a^4}(1+7m)(xy+yz+zx)
\right)u^{\rm BR},\label{eq:vcbrt}\\
\delta V_{\rm C}^{\rm SH1}({\bf r})&=&\frac{2e^2}{4\pi\varepsilon_0}
\left(
-\frac{9\sqrt2}{2a^4}(x^2-y^2)-\frac{3\sqrt2}{a^4}(yz-zx)
\right)(1+7m)u^{\rm SH1},\\
\delta V_{\rm C}^{\rm SH2}({\bf r})&=&\frac{2e^2}{4\pi\varepsilon_0}
\left(
-\frac{3\sqrt6}{2a^4}(x^2+y^2-2z^2)+\frac{\sqrt6}{a^4}(2xy-yz-zx)
\right)(1+7m)u^{\rm SH2}.
\end{eqnarray}
As a result, the coupling constants $a_1^{\rm C}$, $a_2^{\rm C}$, 
$b_1^{\rm C}$ and $b_2^{\rm C}$ are given by
\begin{eqnarray}
a_1^{\rm C}&=&\frac{2e^2}{4\pi\varepsilon_0}\cdot\frac{2\sqrt3}{a^2}\sqrt\frac{\hbar}{2M\omega_{\rm BR}}\left(1-\frac{4r_d^2}{7a^2}(1+7m)\right),\\
a_2^{\rm C}&=&\frac{2e^2}{4\pi\varepsilon_0}\cdot\frac{2\sqrt3}{a^2}\sqrt\frac{\hbar}{2M\omega_{\rm BR}}\left(1+\frac{2r_d^2}{7a^2}(1+7m)\right),\\
b_1^{\rm C}&=&4\sqrt2b_2^{\rm C}=\frac{2e^2}{4\pi\varepsilon_0}\cdot\frac{2\sqrt3}{a^2}\sqrt\frac{\hbar}{2M\omega_{\rm SH}}\frac{4r_d^2}{7a^2}(1+7m).\label{eq:b1b2}
\end{eqnarray}
\end{widetext}
The values of $m$ at $\theta_{15}=84^\circ$ and $82^\circ$ are $m\simeq0.051$ and $0.068$, respectively.
Therefore, $b_1^{\rm C}$ and $b_2^{\rm C}$ at $\theta_{15}=84^\circ$ and $82^\circ$ are
about 40 \% larger than the values at $\theta_{15}=90^\circ$.
The estimated values of $a_1^{\rm C}$, $a_2^{\rm C}$, $b_1^{\rm C}$ and $b_2^{\rm C}$ are shown in Table \ref{table}.
In calculating these values, we use $\frac{2e^2}{4\pi\varepsilon_0 a}=14.4$ eV ($a=2.0$ \AA),
$M=26.561\times10^{-27}$ kg (the mass of $^{16}$O ion), $r_d=0.6$ \AA (radius of Co ion \cite{radius}), $\omega_{\rm BR}=580$ cm$^{-1}$ and $\omega_{\rm SH}=480$ cm$^{-1}$ \cite{frequency}.
In Eqs. (\ref{eq:vcbrt})-(\ref{eq:b1b2}), we show the results up to the first order of $m$.
In calculating the values of $a_1$, $a_2$, $b_1$ and $b_2$ in Table \ref{table},
we use the exact expressions for them with respect to $m$.

Next, we calculate the EPI due to the change of $d$-$p$ transfer integrals.
In the absence of trigonal distortion, we have only to consider the contribution of $t_{pd\pi}$ to $t_{pd}(i,l,m)$.
In contrast, we have also to consider the contribution of $t_{pd\sigma}$ 
in the presence of trigonal distortion, and
the $d$-$p$ transfer integral $t_{pd}(i,l,m)$ is given 
by using the Slater-Koster formula \cite{slater}.
Then, we can obtain $\delta t^\alpha_{pd}(i,l,m)$ 
with the aid of the Harrison's law $t\propto a^{-4}$ \cite{Harrison}.
As a result, EPI can be derived from Eq. (\ref{eq:vt}).
The obtained values of $a_1^{\rm T}$, $a_2^{\rm T}$, $b_1^{\rm T}$ and $b_2^{\rm T}$ are shown in Table \ref{table}.
In calculating these values, we use $t_{pd\sigma}=-1.70$ eV, $t_{pd\pi}=0.785$ eV and $\Delta_{pd}=1.81$ eV according to Ref. \cite{yada1}.

\begin{table}
\begin{tabular}{|l|c|c|c|}
\hline
 $\theta_{15}$ & 90$^\circ$ & 84$^\circ$ & 82$^\circ$ \\
\hline
 $a_1^{\rm C}$ & -1.01 & -0.96 & -0.94 \\
 $a_2^{\rm C}$ & -1.09 & -1.08 & -1.07 \\
 $b_1^{\rm C}$ & 0.0621 & 0.0874 & 0.0971 \\
 $b_2^{\rm C}$ & 0.0110 & 0.0142 & 0.0146 \\
\hline
\end{tabular}
\ \ \ \begin{tabular}{|l|c|c|c|}
\hline
 $\theta_{15}$ & 90$^\circ$ & 84$^\circ$ & 82$^\circ$ \\
\hline
 $a_1^{\rm T}$ & -0.0675 & -0.121 & -0.147 \\
 $a_2^{\rm T}$ & -0.101 & -0.162 & -0.189 \\
 $b_1^{\rm T}$ & 0.0557 & 0.108 & 0.136 \\
 $b_2^{\rm T}$ & 0.0262 & 0.0524 & 0.0629 \\
\hline
\end{tabular}
\caption{Obtained electron-phonon coupling constants 
originating from the change of the Coulomb potential (left panel) and 
the change of $d$-$p$ transfer integrals (right panel) 
for $\theta_{15}=90^\circ$, 84$^\circ$ and 82$^\circ$}\label{table}
\end{table}
\begin{table}
\begin{tabular}{|l|c|c|c|}
\hline
 $\theta_{15}$ & 90$^\circ$ & 84$^\circ$ & 82$^\circ$ \\
\hline
 $a_1$ & -0.215 & -0.218 & -0.218 \\
 $a_2$ & -0.239 & -0.249 & -0.252 \\
 $b_1$ & 0.118 & 0.196 & 0.233 \\
 $b_2$ & 0.0372 & 0.0665 & 0.0775 \\
\hline
\end{tabular}
\caption{Total electron-phonon coupling constants 
for $\theta_{15}=90^\circ$, 84$^\circ$ and 82$^\circ$}\label{table2}
\end{table}

We have calculated the EPI between $t_{2g}$ electrons and zone center phonons ({\bf q}=0).
Hereafter, we neglect the {\bf q}-dependences of the EPI matrix elements 
for simplicity, by expecting that their {\bf q}-dependences
are smeared out after the $\q$-summation.
The dimensionless displacement due to $\alpha$-mode phonon 
in EPI can be expressed as the sum of creation and annihilation operators for 
phonon, $\tilde u_{\bf q}^\alpha=b_{\alpha,{\bf q}}+b^\dag_{\alpha,{\bf q}}$.
Then, the Hamiltonian of EPI between $t_{2g}$ electrons and optical phonons are represented as follows.
\begin{eqnarray}
H_{\rm EPI}=\frac{1}{\sqrt N}\hspace{-2mm}\sum^{BR,SH1,SH2}_{\alpha}\hspace{-1mm}\sum_{\k,\q,\sigma}\hat c^\dag_{{\bf k+q}\sigma}\hat V^\alpha\hat c_{\k\sigma}(b_{\alpha,{\bf q}}+b^\dag_{\alpha,{\bf -q}}),\label{eq:epi}
\end{eqnarray}
where $\hat c_{\k\sigma}=(c_{\k,a_{1g},\sigma}\ c_{\k,e'^1_{g},\sigma}\ c_{\k,e'^2_{g},\sigma})$
is the column vector of annihilation operators for electrons,
and $b_{\alpha,{\bf q}}$ is annihilation operator of $\alpha$-mode phonon.
Then, $\hat V^\alpha$ has the following form:
\begin{eqnarray}
&&\hat V^{\rm BR}=\left(
\begin{array}{ccc}
\ a_1\ &\ 0\ &\ 0\ \\
\ 0\ &\ a_2\ &\ 0\ \\
\ 0\ &\ 0\ &\ a_2\ 
\end{array}
\right),
 \label{eqn:VBR}
\\
\nonumber\\
&&\hat V^{\rm SH1}\hspace{-0.5em}=\left(
\begin{array}{ccc}
0&b_1&0\\
b_1&0&-b_2\\
0&-b_2&0
\end{array}
\right),
\hat V^{\rm SH2}\hspace{-0.5em}=\left(
\begin{array}{ccc}
0&0&-b_1\\
0&b_2&0\\
-b_1&0&-b_2
\end{array}
\right). \nonumber \\
\end{eqnarray}
where $a_i= a_i^C+a_i^T$ and $b_i= b_i^C+b_i^T$ ($i=1,2$).
The estimated values of these coupling constants are shown in Table \ref{table2}.
In estimating $a_1$ and $a_2$, we have considered the screening effect 
by taking account of the chemical potential shift $\delta\mu$:
To conserve the electron number, $\delta\mu$ should satisfy the relation
$\rho_d(0)(\delta\varepsilon_d+\delta\mu)+\rho_p(0)\delta\mu=0$,
where $\rho_d(0)$ and $\rho_p(0)$ are the DOS of $d$-electrons and $p$-electrons at the Fermi level \cite{yada2}.
Then, the shift in the $3d$ level is
$\delta\tilde\varepsilon_d\equiv\delta\varepsilon_d+\delta\mu=\rho_p(0)/[\rho_d(0)+\rho_p(0)]\delta\varepsilon_d$.
Since $\rho_p(0)/[\rho_d(0)+\rho_p(0)]\approx0.2$ in Na$_x$CoO$_2$ \cite{yada1,singh},
both $a_1$ and $a_2$ are reduced to 20\% of their original (unscreened) values.
On the other hand, such a screening effect is absent for shear phonons
since Tr\{$\hat V^{SH}$\} is equal to 0.
In later sections,
we use $b_1=0.20$ and $b_2=0.067$ (eV) for the unhydrated Na$_x$CoO$_2$ ($\theta_{15}=84^\circ$),
and $b_1=0.23$ and $b_2=0.078$ (eV) for the hydrated Na$_x$CoO$_2$ ($\theta_{15}=82^\circ$).
In both cases, we set $a_1=0.22$ and $a_2=0.25$ (eV).

\section{Derivation of Gap Equation}\label{formulation}
\begin{figure}[htbp]
\begin{center}
\includegraphics[scale=0.5]{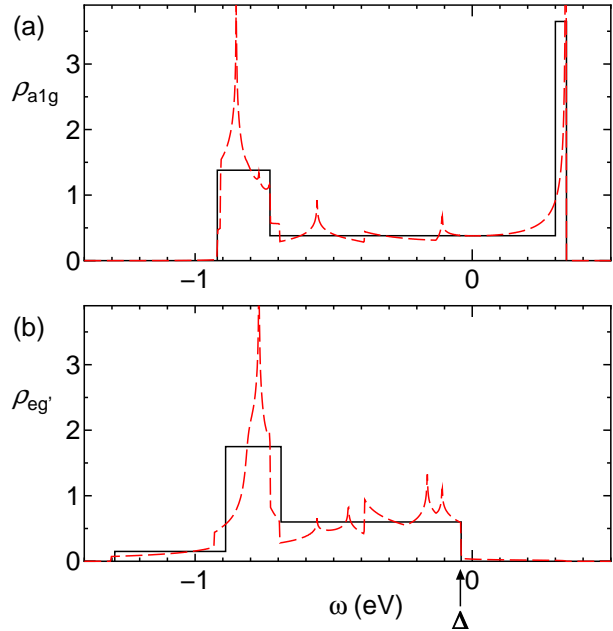}
\caption{(Color online) $d$-electron DOS for the $a_{1g}$-band and the $e_g'$-band per orbit and spin.
Solid line represents the DOS used in the present study (solid line),
and dashed line represents the DOS given by the tight-binding model.
Note that $\rho_{eg'}(\omega)$ in the tight-binding model
takes small values for $\omega>\Delta$,
since the $a_{1g}$-band that forms the large FS around $\Gamma$ point
contains small amount of $e_g'$-orbitals.
}\label{dos}
\end{center}
\end{figure}

In this section, we formulate the model and derive the linearized gap equation.
First, we explain the energy dependence of the DOS in the present model.
In Na$_x$CoO$_2$, there are three electronic bands 
(one $a_{1g}$-band and two $e_g'$-bands) near the Fermi level,
which are composed of $3d$ orbitals of Co ion and $2p$ orbitals 
of O ion in CoO$_2$ layer.
In these bands, the $a_{1g}$-band crosses the Fermi level,
and the $e_g'$-bands are below the Fermi level (valence bands).
Figure \ref{dos} shows the $d$-electron DOS of each band per spin 
that is obtained by the tight-binding model given in Ref. \cite{yada1}.
In the numerical study, we approximate $d$-electron DOS
by the following rectangle function.
\begin{eqnarray}
\rho_{a1g}(\omega)&=&
\left\{
\begin{array}{ccrclc}
1.38&(&-0.92&\le\omega\le&-0.73&)\\
0.38&(&-0.73&\le\omega\le& 0.3&)\\
3.65&(&0.3&\le\omega\le& 0.34&)
\end{array}
\right.,\label{eq:dosa}\\
\rho_{eg'}(\omega)&=&
\left\{
\begin{array}{ccrclc}
0.15&(&\Delta-1.25&\le\omega\le&\Delta-0.85&)\\
1.75&(&\Delta-0.85&\le\omega\le&\Delta-0.65&)\\
0.6&(&\Delta-0.65&\le\omega\le&\Delta&)
\end{array}
\right.,\label{eq:dose}
\end{eqnarray}
where $\Delta \ (<0)$ is the top of the $e_g'$ bands measured from the Fermi level.
The total $d$-electron DOS is $\rho_{a1g}(\omega)+2\rho_{eg'}(\omega)$.
Here, we regard $\Delta$ as variable.
Equations (\ref{eq:dosa}) and (\ref{eq:dose}) are shown in Fig. \ref{dos} (a) and (b), respectively.
Both $\rho_{a1g}(\omega)$ and $\rho_{eg'}(\omega)$
in Eqs. (\ref{eq:dosa}) and (\ref{eq:dose}) are equal to the 
values in the tight binding model at $\omega=0$ and $\omega=\Delta$.
Note that $\int_{-\infty}^\infty \rho_{a1g}(\omega)d\omega
= \int_{-\infty}^\infty \rho_{eg'}(\omega)d\omega = 0.8$
in accord with the band calculation \cite{singh} as well as 
the tight binding model \cite{yada1}.
Here, the number of hole is 0.65,
which corresponds the hole number of Na$_{0.35}$CoO$_2$.

Although LDA band calculation predicts $\Delta>0$ 
(i.e., presence of $e_g'$ hole pockets) \cite{singh},
the relation $\Delta<0$ had been confirmed experimentally
 \cite{yang,hasan,shimojima}.
To solve this discrepancy, many people had studied the strong
correlation effect, which is neglected in the LDA, 
using the Gutzwiller approximation \cite{gutz} and the dynamical
mean field approximation \cite{dmfa1,dmfa2}.
References \cite{gutz,dmfa2} claim that $e_g'$ hole pockets
predicted by the LDA disappears (i.e., $\Delta<0$) due to 
strong Coulomb interaction.
We had also pointed out in Ref. \cite{yada2} that the EPI 
(due to shear phonon) shifts the $e_g'$-bands downwards by -0.05 eV.
In the present paper, we treat $\Delta$ as a model parameter.

\subsection{gap equation}
In Ref. \cite{yada2}, we have studied the linearized gap equation for $s$-wave superconductivity
in the 11-band $d$-$p$ model.
The obtained gap function $\phi_{ll'}(\i\varepsilon_n)$
does not have the off-diagonal components $(l\neq l')$
where $l, l'$ are the indexes of orbital ($l, l'=a_{1g}, e_g'^1, e_g'^2$).
Moreover, gap functions for $e_g'^1$ and $e_g'^2$ are the same.
Then, the linearized gap equation given by Eq. (8) in Ref. \cite{yada2},
which is $3\times3$ matrix equation,
can be reduced to the following $1\times2$ matrix equation:
\begin{eqnarray}
\lambda\hat\phi(\i\varepsilon_n)&=&T\sum_{m}\hat V_{\rm eff}(\i\omega_m)\hat F(\i\varepsilon_n-\i\omega_m),\label{eq:eliash}\\
F_l(\i\varepsilon_n)&=&\frac{1}{N}\sum_{\k,l'}|G_{ll'}(\k,\i\varepsilon_n)|^2\phi_l'(\i\varepsilon_n)
 \nonumber \\
&\cong&\phi_l(\i\varepsilon_n)\frac{1}{N}\sum_{\k}|G_{ll}(\k,\i\varepsilon_n)|^2,\label{eq:sum}
\end{eqnarray}
where $\hat\phi(\i\varepsilon_n)=(\phi_{a1g}(\i\varepsilon_n), \phi_{eg'}(\i\varepsilon_n))$ and $\hat F(\i\varepsilon_n)=(F_{a1g}(\i\varepsilon_n), F_{eg'}(\i\varepsilon_n))$
are column vectors of the gap functions and the linearized anomalous Green functions, respectively,
and $G_{ll'}(\k,\i\varepsilon_n)$ is the normal Green function.
$\varepsilon_n=(2n+1)\pi T$ and $\omega_m=2m\pi T$ are the Matsubara frequencies for fermion and boson, respectively.
In Eq. (\ref{eq:sum}), the eigenvalue of the gap equation, $\lambda$,
increases monotonically as the temperature decreases, and
$T_{\rm c}$ is determined by the condition $\lambda=1$.
In Eq. (\ref{eq:sum}), we ignore the contribution of $l\neq l'$
since the off-diagonal components of Green function are small.
Then, the diagonal component of Green function is approximately given by $G_{ll}(\k,\i\varepsilon_n)=(\i\varepsilon_n-E_{\bf k}^l-\Sigma^l(\k,\i\varepsilon_n))^{-1}$,
where $\Sigma^l(\k,\i\varepsilon_n)$ is the $d$-electron self-energy.
For the effective interaction of Cooper pairs $\hat V_{\rm eff}(\i\omega_m)$,
we consider the EPI given by Eq. (\ref{eq:epi}) and the Coulomb interaction $H_U+H_J$, where
\begin{eqnarray}
H_U&=&\frac{U}{N}\sum_{\bf k,k',q}\sum_l c_{{\bf k},l,\uparrow}^\dag c_{{\bf k'},l,\downarrow}^\dag c_{{\bf k'-q},l,\downarrow}c_{{\bf k+q},l,\uparrow}\\
H_J&=&\frac{J}{N}\sum_{\bf k,k',q}\sum_{l,l'\neq l} c_{{\bf k},l,\uparrow}^\dag c_{{\bf k'},l,\downarrow}^\dag c_{{\bf k'-q},l',\downarrow}c_{{\bf k+q},l',\uparrow}
\end{eqnarray}
where $U$ and $J$ are the on-site Coulomb interaction in the same orbital
and the pair hopping potential between different orbits, respectively.
Hereafter, we set $J=U/10$ according to the first principle calculation 
in Ref. \cite{yada1}.
Since we have assumed that $p$-electrons are noninteracting,
$\hat V_{\rm eff}(\i\omega_m)$ works only for $d$-electrons.
Note that we do not have to consider the Coulomb interaction 
between different orbitals ($U'$)
and the Hund's coupling term ($J_{H}$)
since the off-diagonal components of gap function are absent.
Now, we treat $H_U$ and $H_J$ in the Hartree-Fock approximation,
by extending the theory of Morel and Anderson \cite{morel}.
Then, $\hat V_{\rm eff}(\i\omega_m)$ is given by
\begin{eqnarray}
\hat V_{\rm eff}(\i\omega_m)&=&
\left( 
\begin{array}{cc}
a_1^2 &0 \\
0 &a_2^2 \\
\end{array} 
\right)D_{\rm BR}(\i\omega_m)
+\left( 
\begin{array}{cc}
0 &2b_1^2 \\
b_1^2 &2b_2^2 \\
\end{array} 
\right)D_{\rm SH}(\i\omega_m)\nonumber\\
&&{}-\left( 
\begin{array}{cc}
U &2J \\
J &U+J \\
\end{array}
\right),\label{eq:veff}
\end{eqnarray}
where $D_\alpha(\i\omega_m)=2\omega_{\alpha}/(\omega_\alpha^2+\omega_m^2)$ is the phonon Green function for $\alpha$-mode.
$\omega_{\alpha}$ is the Debye frequency of $\alpha$-mode phonon.
The first two terms in Eq. (\ref{eq:veff}) represent attractive forces 
due to EPI, and the last term
represents the repulsive force due to the Coulomb interaction.
We see that attractive force induced by breathing phonon, 
which corresponds to the first term in Eq. (\ref{eq:veff}), 
is reduced by $U$ (or $U+J$).
On the other hand, attractive force due to shear phonon
is reduced by $J$ (or $2J$).

Here, we calculate $F_l(\i\varepsilon_n)$ in Eq. (\ref{eq:sum}).
Instead of summing \k in the right-hand side, we perform the energy integration as
\begin{eqnarray}
\frac{1}{N}\sum_{\k}|G_{ll}(\k,\i\varepsilon_n)|^2=\int_{-\infty}^\infty\frac{z_l(\omega)\rho_l(\omega)}{\omega^2+\varepsilon_n^2}d\omega,\label{eq:g2}
\end{eqnarray}
where $\rho_l(\omega)$ is the $d$-electron DOS defined in Eqs. (\ref{eq:dosa})-(\ref{eq:dose}),
and $z_l(\omega)$ is the renormalization factor given by the self-energy.
The renormalization factor due to EPI at $\omega=0$ is given by
$z_l(0)=\left(1-\left.\frac{\partial\Sigma^{\rm BR}_l(\omega)}{\partial\omega}\right|_{\omega=0}-\left.\frac{\partial\Sigma^{\rm SH}_l(\omega)}{\partial\omega}\right|_{\omega=0}\right)^{-1}$,
where $\Sigma^{\rm BR}_l(\omega)$ and $\Sigma^{\rm SH}_l(\omega)$ are the self-energies given by breathing and shear phonons, respectively.
Since the renormalization due to the self-energy 
takes place only for $|\omega|<\omega_{\alpha}$ \cite{yada2},
we approximate $z_l(\omega)$ as 
\begin{widetext}
\begin{eqnarray}
z_l(\omega)=
\left\{
\begin{array}{cl}
\left(1-\left.\frac{\partial\Sigma^{\rm BR}_l(\omega)}{\partial\omega}\right|_{\omega=0}-\left.\frac{\partial\Sigma^{\rm SH}_l(\omega)}{\partial\omega}\right|_{\omega=0}\right)^{-1}&(|\omega|<\omega_{\rm SH})\\
\left(1-\left.\frac{\partial\Sigma^{\rm BR}_l(\omega)}{\partial\omega}\right|_{\omega=0}\right)^{-1}&(\omega_{\rm SH}<|\omega|<\omega_{\rm BR})\\
1&(\omega_{\rm BR}<|\omega|)
\end{array}
\right. \label{eq:z}
\end{eqnarray}
\end{widetext}
Note that $T_{\rm c}$ is reduced by $z_l(\omega)$.
In the one-loop approximation \cite{yada2}, $\Sigma^{\rm BR}_l(\omega)$ and $\Sigma^{\rm SH}_l(\omega)$ are given by
\begin{eqnarray}
\Sigma^{\rm BR}_{a1g}(i\varepsilon_n)&=&T\sum_{m}a_1^2G_{a1g}(i\varepsilon_n-i\omega_m)D_{\rm BR}(i\omega_m),\\
\Sigma^{\rm BR}_{eg}(i\varepsilon_n)&=&T\sum_{m}a_2^2G_{eg}(i\varepsilon_n-i\omega_m)D_{\rm BR}(i\omega_m),
\end{eqnarray}
\begin{eqnarray}
\Sigma^{\rm SH}_{a1g}(i\varepsilon_n)&=&T\sum_{m}2b_1^2G_{eg}(i\varepsilon_n-i\omega_m)D_{\rm SH}(i\omega_m),\\
\Sigma^{\rm SH}_{eg}(i\varepsilon_n)&=&T\sum_{m}\left\{b_1^2G_{a1g}(i\varepsilon_n-i\omega_m) \right. 
 \nonumber \\
& &\left.+2b_2^2G_{eg}(i\varepsilon_n-i\omega_m)\right\}D_{\rm SH}(i\omega_m).
\end{eqnarray}
In deriving above equations, we have put $D_{\alpha}(\i\omega_m)=\frac{2}{\omega_{\alpha}}\theta(\omega_{\alpha}-|\omega_m|)$ for simplicity.
Then, the differential coefficient of self-energy at $\omega=0$ is expressed in the following form.
\begin{widetext}
\begin{eqnarray}
-\frac{\partial}{\partial\omega}\left(
\begin{array}{c}
\Sigma^{\rm BR}_{a1g}(\omega)\\
\Sigma^{\rm BR}_{eg'}(\omega)
\end{array}
\right)\Bigg|_{\omega=0}
&=&
\frac{2}{\pi}\int_{-\infty}^\infty
\left(
\begin{array}{cc}
a_1^2&0\\
0&a_2^2\\
\end{array}
\right)
\left(
\begin{array}{c}
\rho_{a1g}(\varepsilon)\\
\rho_{eg'}(\varepsilon)\\
\end{array}
\right)
\frac{1}{\varepsilon^2+\omega_{\rm BR}^2}d\varepsilon,\label{eq:zw1}
\\
-\frac{\partial}{\partial\omega}\left(
\begin{array}{c}
\Sigma^{\rm SH}_{a1g}(\omega)\\
\Sigma^{\rm SH}_{eg'}(\omega)
\end{array}
\right)\Bigg|_{\omega=0}&=&\frac{2}{\pi}\int_{-\infty}^\infty
\left(
\begin{array}{cc}
0&2b_1^2\\
b_1^2&2b_2^2\\
\end{array}
\right)
\left(
\begin{array}{c}
\rho_{a1g}(\varepsilon)\\
\rho_{eg'}(\varepsilon)\\
\end{array}
\right)
\frac{1}{\varepsilon^2+\omega_{\rm SH}^2}d\varepsilon
.\label{eq:zw}
\end{eqnarray}
\end{widetext}
In the next section, we solve Eqs. (\ref{eq:eliash})-(\ref{eq:sum}) 
self-consistently using Eqs. (\ref{eq:g2})-(\ref{eq:z}),
and obtain $T_{\rm c}$ that is determined by the condition
$\lambda=1$ in Eq. (\ref{eq:eliash}).

\section{numerical result}\label{result}

Here, we perform numerical study:
In calculating $T_{\rm c}$,
important model parameters are $a_1$, $b_1$ and $\Delta$:
$a_1$ ($b_1$) represents the EPI due to breathing mode (shear mode)
in the $a_{1g}$-orbital (between $a_{1g}$- and $e_g'$-orbitals),
and $\Delta$ is the energy of  the top of the $e_g'$ bands measured from the Fermi level in Fig. \ref{dos} (b).
Note that $\Delta<0$ experimentally \cite{yang,hasan,shimojima}.
The values of $a_1$ and $b_1$ obtained in the present paper
is slightly larger than those used in Ref. \cite{yada2}.
Therefore, the mass enhancement factors in the present model
given by Eqs. (\ref{eq:zw1}) and (\ref{eq:zw})
are slightly larger than those in Fig. 2 of Ref. \cite{yada2}:
In the present model, $z_{a1g}^{-1}(0)$
is around 2 for $\Delta\sim-0.02$ eV, and it decreases moderately
as $\Delta$ decreases.

Then, $T_{\rm c}$ is obtained 
by solving the set of gap equations (\ref{eq:eliash})-(\ref{eq:sum}).
In the absence of Coulomb interaction,
the obtained $T_{\rm c}$-$\Delta$ phase diagram
is qualitatively similar to Fig. 4 in Ref. \cite{yada2},
where obtained $T_{\rm c}$ is about 10 times higher than 
the experimental $T_{\rm c}$.
In this section, we solve the gap equation in the presence of
strong Coulomb interaction between $d$-electrons.
We will show that the obtained $T_{\rm c}$ is reduced 
to be comparable with experimental $T_{\rm c}$
by assuming reasonable values of $U$ and $J$.
In the presence of shear phonons ($b_1\ne0$), 
gap function for $e_g'$ orbital $\phi_{eg'}(\i\varepsilon_n)$ is finite 
even if the $e_g'$ bands sink below the Fermi level ($\Delta<0$).
Due to this interband transition of Cooper pairs,
which we call the valence band SK effect,
$s$-wave $T_{\rm c}$ is considerably enhanced.

\begin{figure}[htbp]
\begin{center}
\includegraphics[scale=0.45]{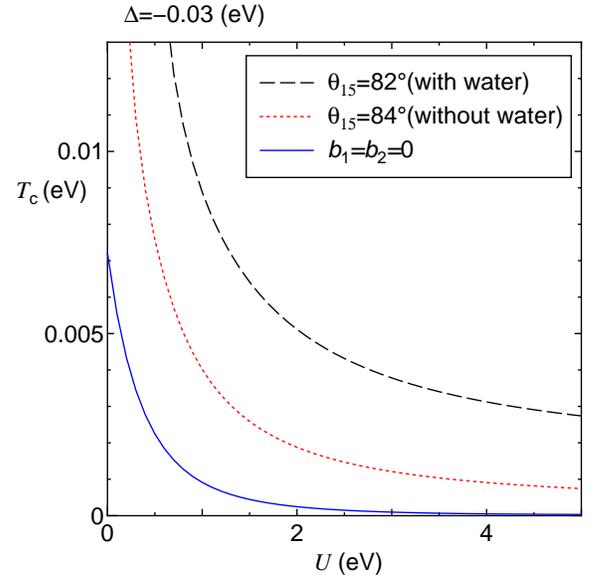}
\caption{(Color online) $U$-dependence of $T_{\rm c}$ for $\Delta=0.03$ eV.}\label{fig1}
\end{center}
\end{figure}

In Fig. \ref{fig1}, 
we show $U$-dependence of $T_{\rm c}$ at $\Delta=-0.03$ eV
for unhydrate ($b_1=0.2$ eV and $b_1=0.067$ eV) and hydrate 
($b_1=0.23$ eV and $b_2=0.78$ eV) systems, respectively.
We see that $T_{\rm c}$ decrease monotonically with $U$
since the attractive force due to phonons is reduced by $U$.
In the absence of shear phonons ($b_1=b_2=0$),
$T_{\rm c}$ is independent of $\Delta$.
Experimental $T_{\rm c}\sim4\times10^{-4}$ eV is realized
at $U\sim1.75$ eV, whereas $T_{\rm c}$ is almost zero for $U\ge4$ eV.
In the presence of both breathing and shear phonons, obtained $T_{\rm c}$
is considerably enlarged 
owing to the valence band SK effect induced by shear phonons,
which is represented by $b_1^2$ in Eq. (\ref{eq:veff}).
For $U\sim4$ eV, obtained $T_{\rm c}$ for hydrate system 
is more than three times larger than that for unhydrate one.
In \S \ref{sec:retardation}, we explain the reason why
SK effect is strong against large Coulomb interaction.

\begin{figure}[htbp]
\begin{center}
\includegraphics[scale=0.49]{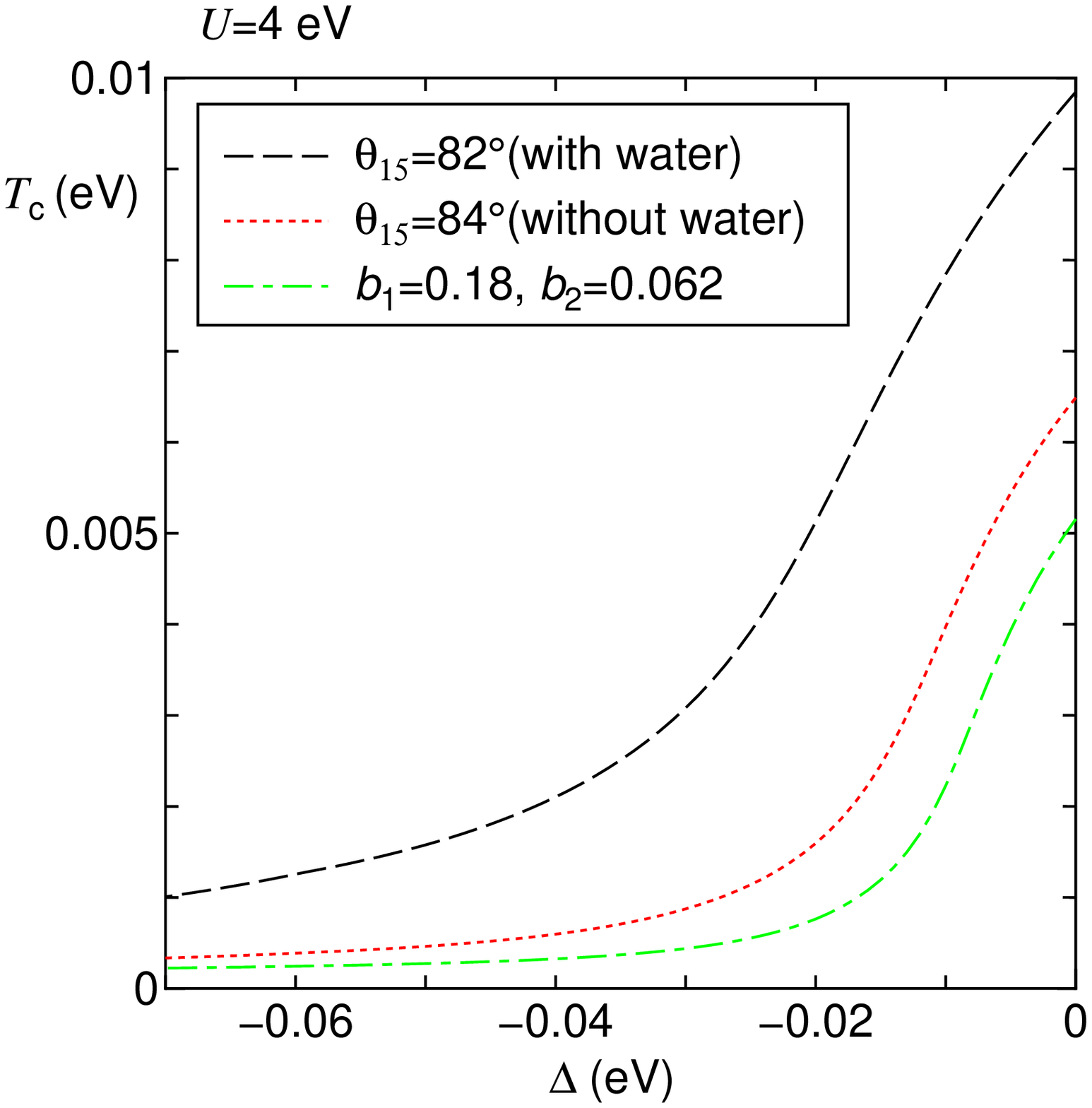}
\includegraphics[scale=0.49]{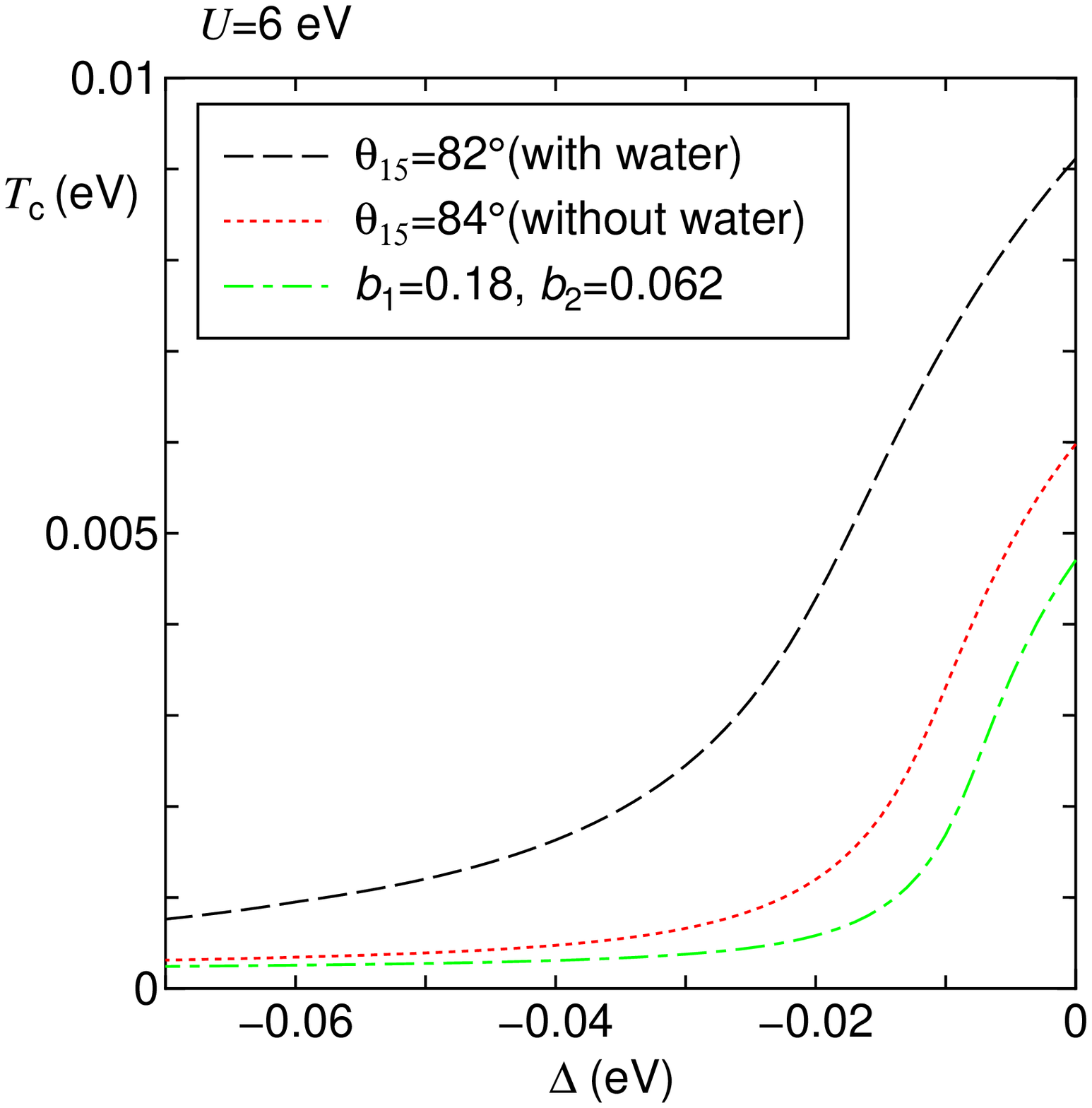}
\caption{(Color online) $\Delta$-dependence of $T_{\rm c}$ at $U$=4 eV (top) and $U$=6 eV (bottom).
The dashed and dotted line represent $T_{\rm c}$ for hydrated Na$_x$CoO$_2\cdot y$H$_2$O ($b_1=0.223, b_2=0.0775$) and unhydrated Na$_x$CoO$_2$
($b_1=0.196, b_2=0.0665$), respectively.
}\label{fig2}
\end{center}
\end{figure}

In Fig. \ref{fig2}, we show the $\Delta$-dependence of $T_{\rm c}$ at $U=4$ and 6 eV for unhydrate and hydrate systems, respectively.
Note that the cluster calculation using quantum chemical ab-initio methods suggests $U=4.1\sim4.8$ eV \cite{landron}.
In both cases, $T_{\rm c}$ is almost zero if we drop shear phonons ($b_1=b_2=0$).
In the presence of shear phonons, on the other hand,
obtained $T_{\rm c}$ is comparable to or higher than 
experimental $T_{\rm c}$ due to the valence band SK effect 
for $|\Delta|\lesssim \omega_{\rm SH}$.
In Fig. \ref{fig2}, $T_{\rm c}$ increases monotonically with 
increasing $\Delta\ (<0)$ due to SK effect, in particular 
for $|\Delta|<\omega_{\rm SH}$.
ARPES study in hydrated Na$_x$CoO$_2\cdot y$H$_2$O \cite{shimojima} suggests that $\Delta\sim-0.03$ eV,
which is closer to the Fermi level than the Debye frequency of shear phonon
$\omega_{\rm SH}$=480 cm$^{-1}\sim$0.06 eV.
On the other hand, ARPES studies in unhydrated Na$_x$CoO$_2$ \cite{yang,hasan} suggest $\Delta\sim-0.1$ eV.
Therefore, our results are consistent with these experimental results.

The theoretical value of $T_{\rm c}$ in Fig. \ref{fig2} for 
$|\Delta|\sim0.02-0.03$ eV is much larger than experimental $T_{\rm c}$,
even for parameters of ``without water''.
Therefore, we also calculate $T_{\rm c}$ for smaller values of 
$b_1$ and $b_2$ in Fig. \ref{fig2};
$b_1=0.18$ eV and $b_2=0.062$ eV.
The obtained $T_{\rm c}$ for $|\Delta|\sim0.02-0.03$ eV is comparable 
with experimental value $T_{\rm c}=5{\rm K} \approx 0.04$ eV.

In Appendix, we analytically study the transition temperature
by taking account of the valence band SK effect.
The obtained expression for $T_{\rm c}$ is given in Eq. (\ref{eq:Tc}),
where the effective coupling constant $\lambda_{\rm eff}$
is given in Eq. (\ref{eq:lambda}).
Since $\lambda_2, \lambda_3 \propto b_1$,
the second term in Eq. (\ref{eq:lambda}) represents the contribution
due to the valence band SK effect. 
This term increases as $\Delta$ approaches zero
since $\lambda_\Delta$ is a decrease function of $\Delta\ (<0)$.

Finally, we summarize the results of this section.
Owing to the valence band SK effect,
the obtained $s$-wave $T_{\rm c}$ is relatively high
even if we take account of the realistic Coulomb repulsion $U=4\sim6$ eV.
Moreover, not only $\Delta$ approaches zero 
but also EPI for shear phonons ($b_1$ and $b_2$) increase 
due to the change of crystal structure by water intercalation,
as explained in \S \ref{epi2}.
For this reason, $s$-wave superconductivity is realized against
strong Coulomb interaction in hydrated Na$_x$CoO$_2\cdot y$H$_2$O.

\section{discussion}\label{discussion}
\subsection{Why SK mechanism can overcome strong Coulomb interaction?}\label{sec:retardation}

In \S \ref{result}, we have shown that
$T_{\rm c}$ becomes 0 at relatively small $U$($\sim4$ eV) when we consider only breathing phonon,
while large $T_{\rm c}$ is realized even for $U\ge4$ eV when we consider shear phonons as well as breathing phonon.
This result indicates that the SK mechanism is suppressed by 
the Coulomb interactions only slightly.
In fact,
the attractive force due to shear phonons is reduced by $J\sim U/10$,
which means that the reduction of the SK mechanism due to Coulomb interaction
is very small.
Moreover, $J$ is prominently reduced by the retardation effect,
as we will explain below:

In a single band model, the retardation effect 
was discussed by Morel and Anderson \cite{morel}.
In Appendix, we extend their theory to the multi-band model for Na$_x$CoO$_2$.
The renormalized Coulomb interactions are obtained by solving 
set of equations (\ref{eq1})-(\ref{eq4}).
Here, we approximate the DOS for the $a_{1g}$- and $e_g'$-bands as 
$\rho_{a1g}(\omega)= \rho_{a1g}\theta(W-|\omega|)$ and
$\rho_{eg'}(\omega)= \rho_{eg'}\theta(\Delta-\omega)\theta(W+\omega)$
[see Fig. \ref{DOS2} in appendix].
Here, 2$W$ is the bandwidth of the $a_{1g}$-band.
In a realistic parameter regime $U\gg J$, these equations can be solved easily.
The obtained results are
\begin{eqnarray}
U^*_{a1g}&\approx&U/(1+AU),\label{eq:ua1g}\\
U^*_{eg'}&\approx&U/(1+BU),\\
J^*_{a1g,eg'}&\approx&J/(1+AU)(1+BU),\label{eq:ja1g}\\
J^*_{eg',eg'}&\approx&J/(1+BU)^2,\label{eq:jeg}
\end{eqnarray}
where $U^*_{a1g}$ ($U^*_{eg'}$) is renormalized Coulomb interaction between $a_{1g}$ electrons ($e_g'$ electrons),
$J^*_{a1g,eg'}$ ($J^*_{eg',eg'}$) is renormalized pair hopping between $a_{1g}$ and $e_g'$ orbitals ($e_g'$ orbitals).
$A=\rho_{a1g}\ln(W/\omega_D)$ and $B=\frac12 \rho_{eg'}\ln(W/\omega_D)$ 
are the renormalization factors.
In the present model, $A\approx1.2$ and $B\approx1.0$ \cite{yada3}.
(Note that $\rho_{a1g}U^*_{a1g}$ ($\rho_{eg'}U^*_{eg'}$) corresponds to
Anderson-Morel pseudopotential $\mu$ for a single band model.)
Since the renormalization factor for $J$ in Eqs. (\ref{eq:ja1g}) and (\ref{eq:jeg}) is square of that for $U$, $J^*$ becomes very small.
In the present model,
$U^*_{a1g}/U\approx 1/7$ and $J^*_{a1g,eg'}/J\approx1/42$ at $U\approx 5$ eV.
For this reason, the effect of the Coulomb interactions on the SK mechanism becomes remarkably small.
Therefore, in the case of $|\Delta|\ll\omega_{\rm D}$,
$s$-wave superconducting state is realized owing to the SK effect
even in the presence of strong Coulomb interaction.

\subsection{Isotope effect on $T_{\rm c}$}

Recently, Yokoi et al. studied the isotope effect 
on $T_{\rm c}$ in Na$_x$CoO$_2\cdot y$H$_2$O by substituting
$^{16}$O atoms with $^{18}$O atoms \cite{yokoi_isotope}.
They found that the isotope effect coefficient $\alpha$
($T_{\rm c}\propto M^{-\alpha}$, where $M$ is the mass of O ion)
is considerably smaller than the simple BCS value 0.5.
Here, we study the isotope effect on $T_{\rm c}$ in the present model,
and explain that the value of $\alpha$ approaches zero
because of the strong Coulomb interaction.

According to the BCS theory, $T_{\rm c}$ in a single band model is 
$T_{\rm c}= 1.13\omega_D e^{-1/\lambda^*}$ and
$\lambda^*= 2N(0)a^2/\omega_D -\mu^*$, where
$N(0)$ is the density of states, $a$ is the EPI, and 
$\mu^*=N(0)U/(1+N(0)U\ln(W/\omega_D)$ is the Anderson-Morel pseudo potential.
In the case of $\mu^*=0$ ($U=0$), $2N(0)a^2/\omega_D$ is independent of $M$
since $a\propto (M\omega_D)^{-1/2} \propto M^{-1/4}$.
Then, $T_{\rm c}\propto\omega_{\rm D}\propto M^{-1/2}$ (i.e., $\alpha=0.5$).
However, $\alpha$ becomes smaller than 0.5 in the case of $U>0$, since 
$\mu^*$ decreases as $M \ (\propto \omega_D^{-2})$ increases.
In fact, $\alpha\sim0$ in Ru and Zr, 
both of which are 4$d$-electron $s$-wave superconductors.
Since both $N(0)U$ and $\omega_D/W$ are relatively large 
in Na$_x$CoO$_2\cdot y$H$_2$O,
the value of $\alpha$ is expected to be much smaller than 0.5.

\begin{figure}[htbp]
\begin{center}
\includegraphics[scale=0.5]{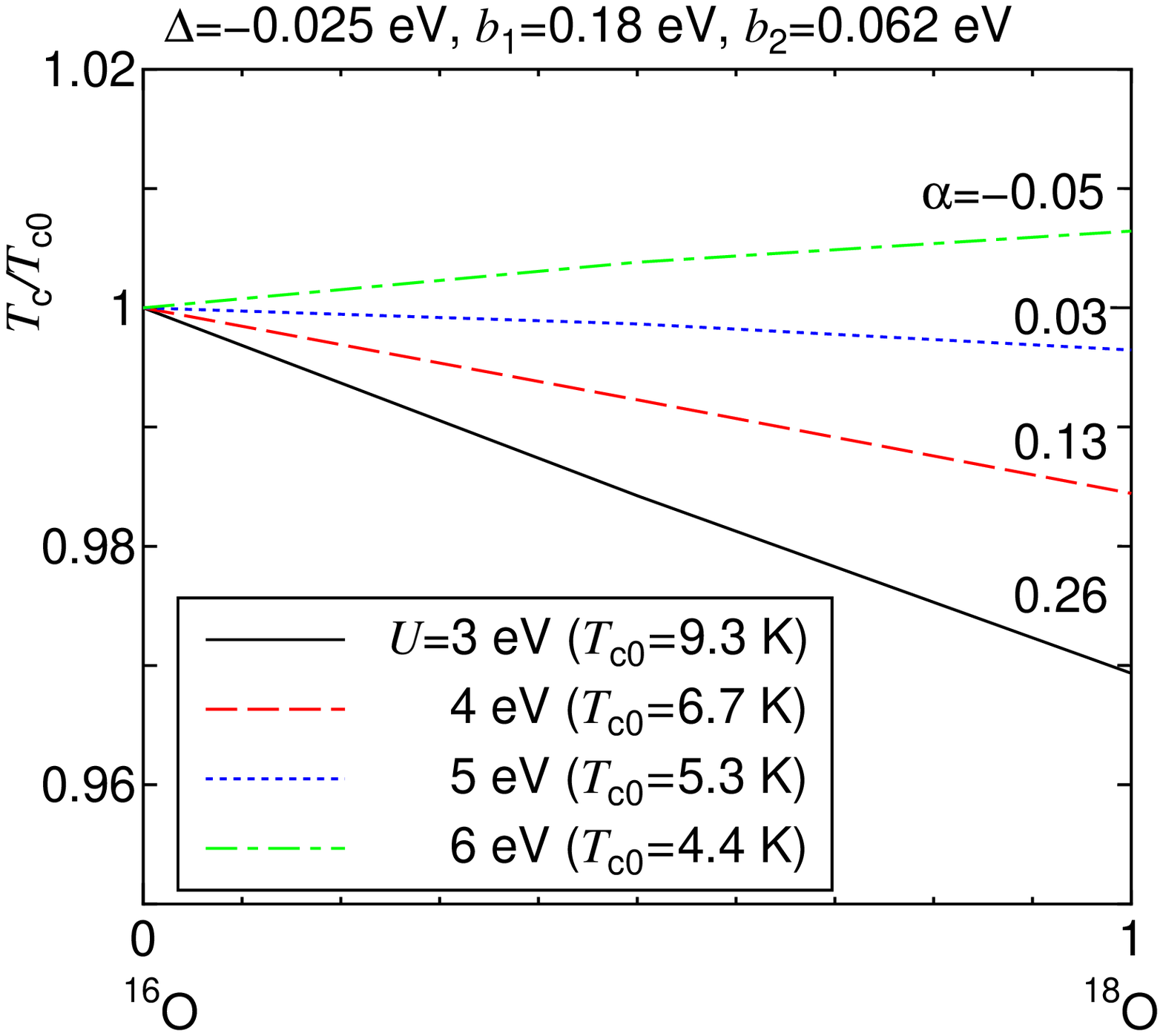}
\includegraphics[scale=0.5]{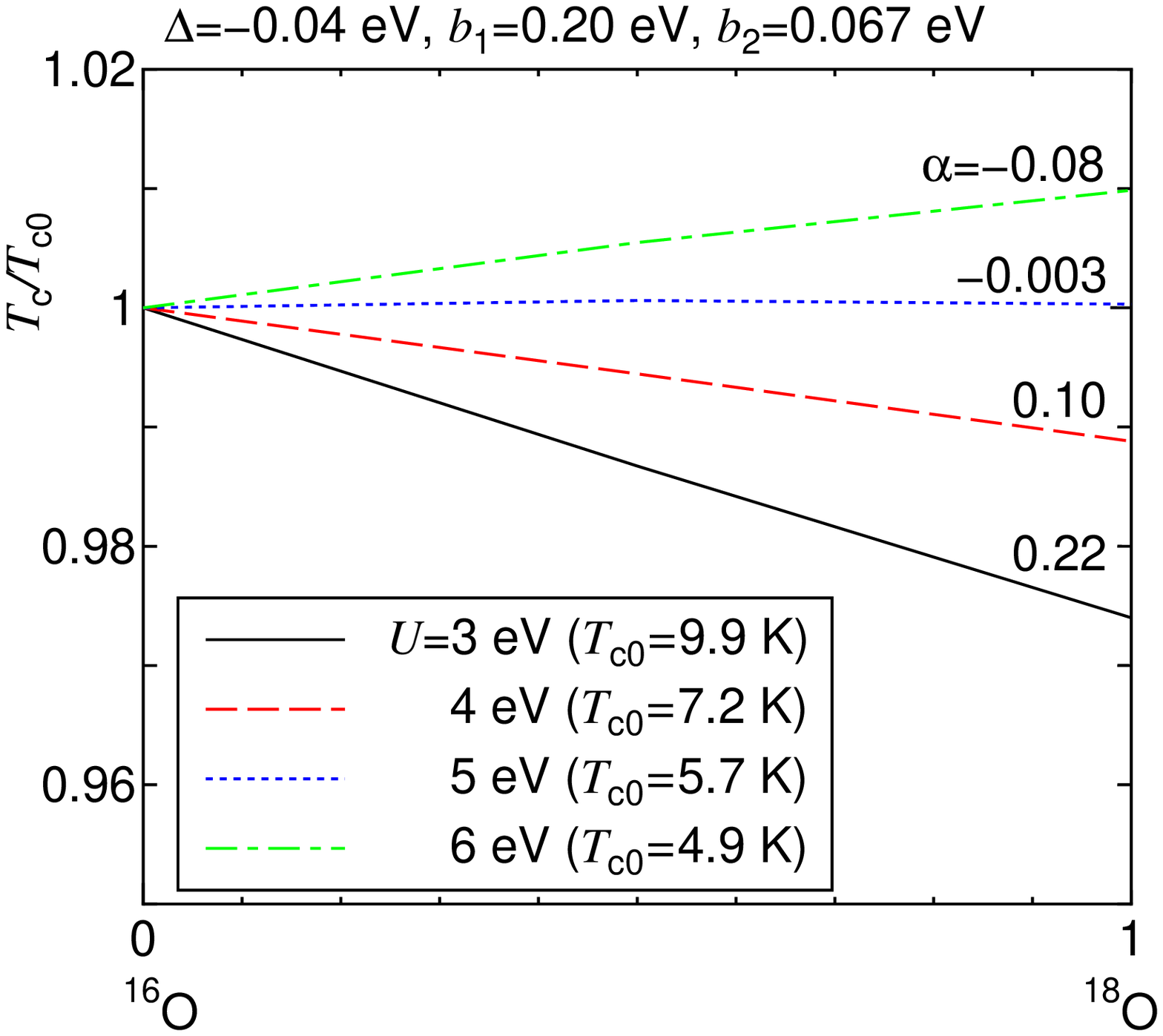}
\caption{(Color online) Isotope effect on $T_{\rm c}$ for at $\Delta=-0.025$ eV, $b_1=0.18$ eV, $b_2=0.062$ eV (top),
and $\Delta=-0.04$ eV, $b_1=0.20$ eV, $b_2=0.067$ eV (bottom).
The lateral axis shows the rate of $^{18}$O ions and
the longitudinal axis shows $T_{\rm c}/T_{\rm c0}$ where $T_{\rm c0}$ is $T_{\rm c}$ for $^{16}$O ion.
Note that $\alpha=-\partial\ln T_{\rm c}/\partial\ln M$ is the 
isotope effect coefficient.}\label{fig4}
\end{center}
\end{figure}

Figure \ref{fig4} show the obtained isotope effect on $T_{\rm c}$
at $\Delta=-0.025$ eV and $\Delta=-0.04$ eV, by assuming
$\omega_\a \propto M^{-1/2}$ and 
$a_1, a_2, b_1, b_2 \propto M^{-1/4}$.
For $b_1$ and $b_2$, we use the smaller values than the estimated ones in Table \ref{table}
to reproduce the experimental $T_{\rm c}(\sim4.5$ K).
We use $b_1=0.18$ eV, $b_2=0.062$ eV for $\Delta=-0.025$ eV,
and $b_1=0.20$ eV, $b_2=0.067$ eV for $\Delta=-0.04$ eV.
(In both cases, we put $a_1=0.22$ eV and $a_2=0.25$ eV, 
which are same with the estimated values in Table \ref{table2}.)
For both parameters,
the reduction of $T_{\rm c}$ due to isotope effect, $\Delta T_{\rm c}$, decreases with $U$.
We see that $\Delta T_{\rm c}$ becomes almost 0 at $U\sim5$ eV,
and it becomes positive at $U=6$ eV.
Therefore, $\Delta T_{\rm c}$ in Na$_x$CoO$_2$ may be
too small to probe it experimentally.
At $U=3$ eV, the obtained $\alpha$ is about half of the BCS value ($\alpha=0.5$).
$\alpha$ becomes almost 0 at $U=5$ eV, and it becomes negative at $U=6$ eV
(inverse isotope effect).
Therefore, $\alpha$ approaches zero in Na$_x$CoO$_2\cdot y$H$_2$
due to the strong Coulomb interaction,
even if the $s$-wave superconductivity is caused by EPI.

\subsection{Origin of the anisotropy of the superconducting gap}

In this paper, we have studied the $s$-wave superconducting state
due to EPI in Na$_x$CoO$_2\cdot y$H$_2$O.
However, the nuclear relaxation ratio $1/T_1$ is proportional 
to $T^3$ below $T_{\rm c}$, which suggests that the 
nodal superconducting state is realized \cite{zheng_nqr}.
Presence of gap anisotropy is also indicated by the specific heat
measurements below $T_{\rm c}$ \cite{Cv1,Cv2}.
In the present calculation, we have neglected the 
${\bf k}$-dependence of the superconducting gap for simplicity.
Here, we discuss the possibility of realizing an anisotropic 
$s$-wave state in Na$_x$CoO$_2\cdot y$H$_2$O, resulting from the coexistence
of the strong EPI and the AF fluctuations.

In (Y,Lu)Ni$_2$B$_2$C, strongly anisotropic superconducting state
is realized below $T_{\rm c}\sim 15$ K \cite{Izawa,watanabe}.
The superconducting gap becomes isotropic
by introducing small amount of impurities, whereas
$T_{\rm c}$ is almost unchanged \cite{Nohara}.
They are hallmarks of anisotropic $s$-wave superconducting state.
According to Ref. \cite{Izawa}, the ratio of gap anisotropy reaches 
$\Delta_{\rm max}/\Delta_{\rm min}\sim10$ in a clean sample.
Recently, one of the authors of this paper had studied the mechanism
of this anisotropic $s$-wave superconductivity \cite{kontani}:
By solving the strong coupling Eliashberg equation,
he found that the $s$-wave superconducting gap due to the EPI
can be strongly anisotropic even in a single FS model,
if strong AF fluctuations exist.
In this case, pairs of gap minima appear at points on the FS
which are connected by the nesting vector {\bf Q}.
In the normal state of Na$_x$CoO$_2\cdot y$H$_2$O,
prominent AF fluctuations had been observed by
$1/T_1$ measurements \cite{ishida2,zheng_nqr}.
According to Ref. \cite{ishida2}, Na$_x$CoO$_2\cdot y$H$_2$O locates
in the close vicinity of the AF quantum critical point.
Recent neutron diffraction measurement reports that 
the wavevector of the AF fluctuations is 
${\bf Q}_2=\overrightarrow{\Gamma {\rm M}}=(0, 2\pi/\sqrt3)$
and $(\pi,\pm\pi/\sqrt{3})$ \cite{moyoshi-private}.

\begin{figure}[htbp]
\begin{center}
\includegraphics[scale=0.4]{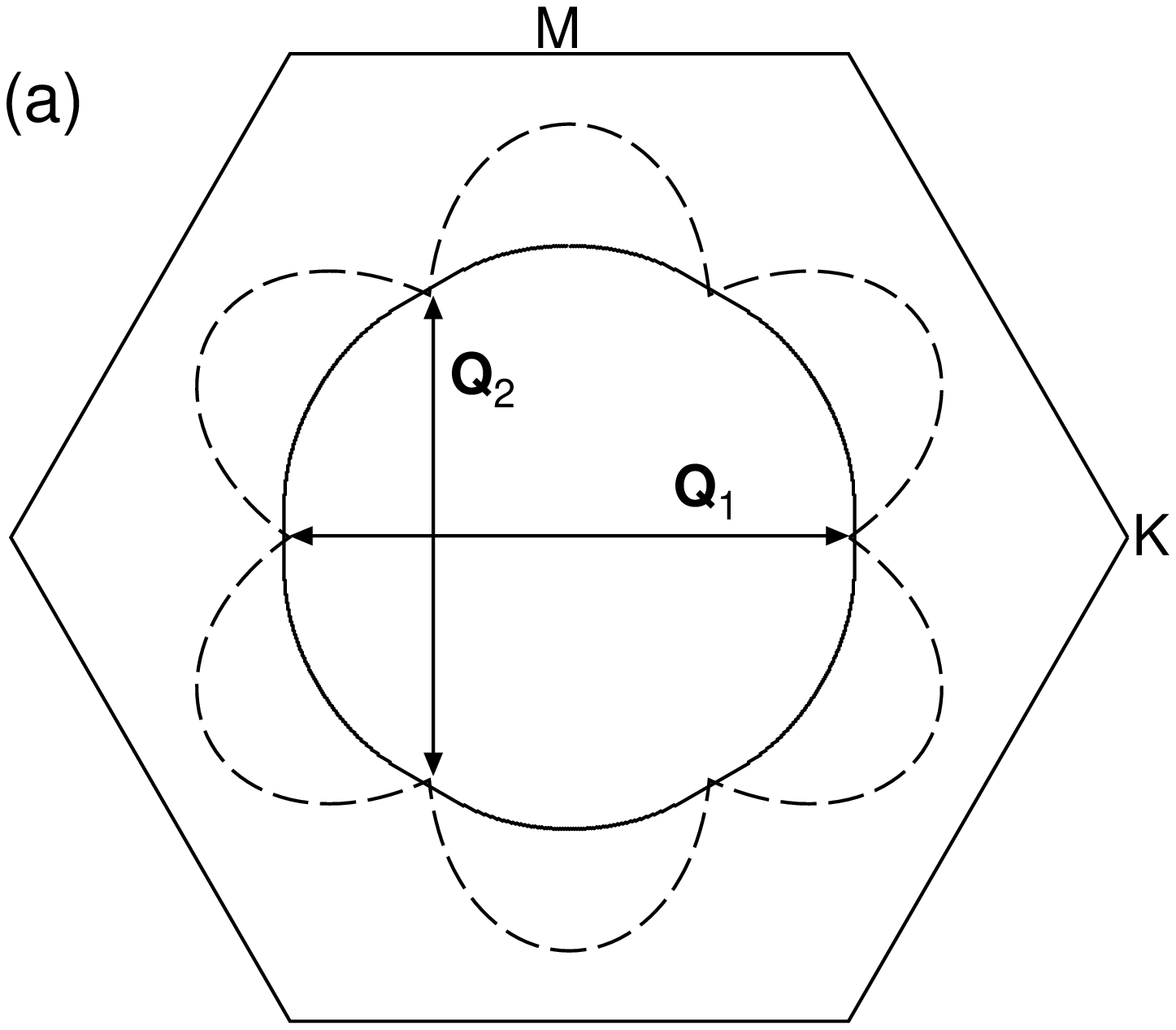}
\includegraphics[scale=0.4]{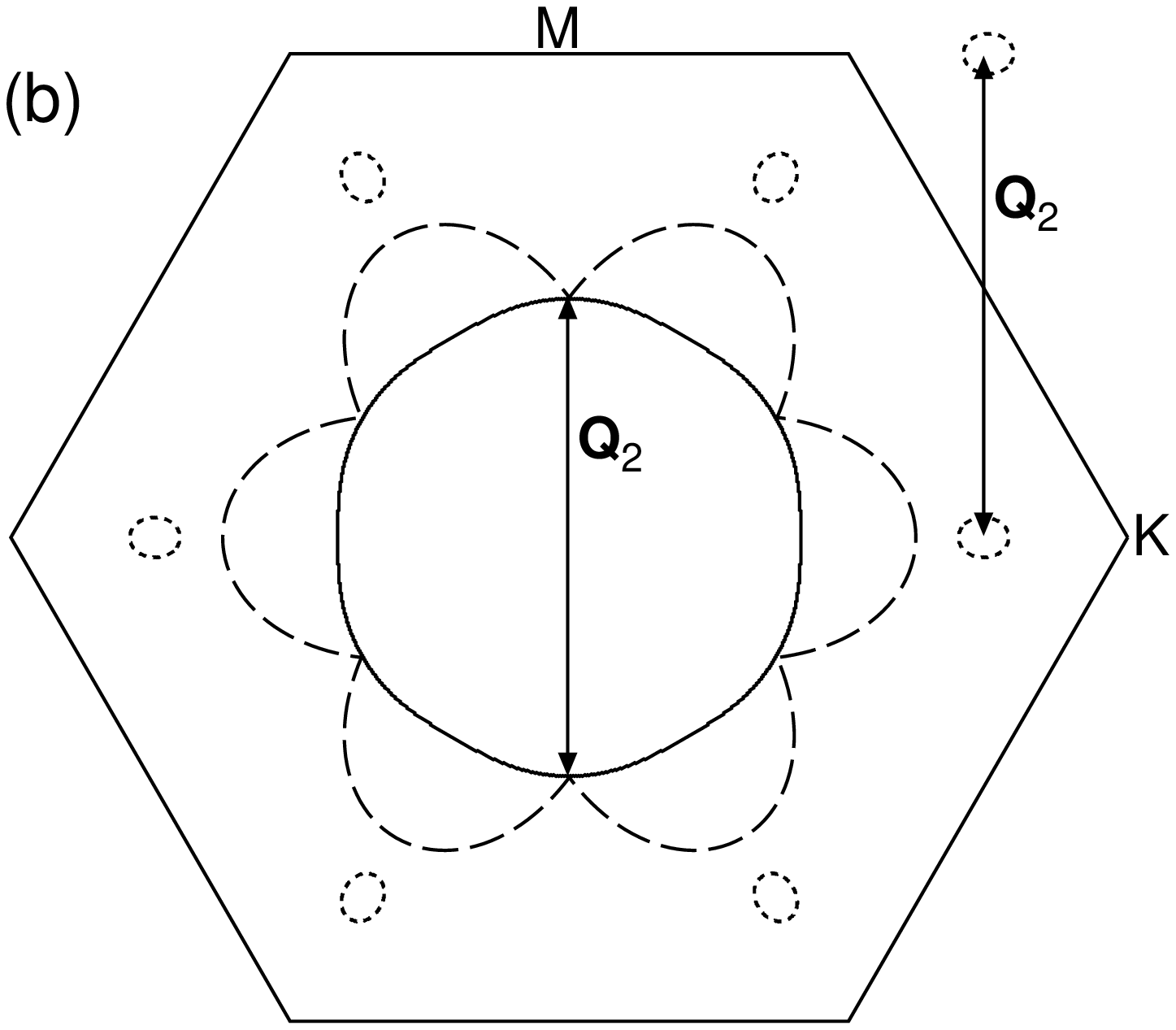}
\caption{Expected anisotropic $s$-wave superconducting gap 
for (a) $n_h\sim0.65$ and (b) $n_h\sim0.5$,
where $n_h$ is the hole number of $t_{2g}$ orbital.
The dashed lines represent the magnitude of gap function,
whose sign is positive everywhere .
Tops of the $e_g'$-bands, which are just below the Fermi level,
are shown by dotted lines in (b).
}\label{fig:gap}
\end{center}
\end{figure}

According to the study using the FLEX approximation \cite{yada1}, 
there are two candidates for the nesting vectors in Na$_x$CoO$_2$ for $\Delta<0$.
One is ${\bf Q}_1=\overrightarrow{\Gamma {\rm K}}=(4\pi/3, 0)$ 
and $(\pm 2\pi/3,2\pi/\sqrt{3})$ that originates from the 
$a_{1g}$ large FS around $\Gamma$-point.
The another one is ${\bf Q}_2$
that originates from the nesting of $e_g'$ hole pockets near K-points.
Note that the AF fluctuations with ${\bf Q}_2$ appear when the top of 
the $e_g'$ bands is very close to Fermi level \cite{yada1}.
When the hole number is $0.65$, which corresponds to the 
hole number of Na$_{0.35}$CoO$_2\cdot 1.3$H$_2$O without oxonium ion,
the resultant minimum gap points are shown in Fig \ref{fig:gap} (a),
when the wavenumber of AF fluctuations is either ${\bf Q}_1$ or ${\bf Q}_2$.
Note that the sign of the gap function is the same everywhere.
On the other hand, when the hole number is about $0.5$,
which can be realized in the presence of oxonium ion 
as a substitute for water molecule \cite{oxonium},
the resultant minimum gap points are shown in Fig \ref{fig:gap} (b)
when the wavenumber of AF fluctuations is ${\bf Q}_2$.
In both cases, anisotropic $s$-wave superconductivity 
is realized in the $e_g'$-band FS in Na$_x$CoO$_2\cdot y$H$_2$O.

In Ref. \cite{kontani}, we have shown that the relation $1/T_1\propto T^3$ 
below $T_{\rm c}$ can be realized in the anisotropic $s$-wave state
under the influence of strong AF fluctuations.
Moreover, we have recently verified that the two-gap type
specific heat observed in Refs. \cite{Cv1,Cv2} can be 
realized in this anisotropic $s$-wave state,
by considering only the $a_{1g}$ FS \cite{kontani2}.
It is an important future problem to reproduce the 
anisotropic $s$-wave state microscopically
based on the $d$-$p$ Holstein Hubbard model for Na$_x$CoO$_2$.

\section{summary}\label{summary}
In Na$_x$CoO$_2$, existence of the strong EPI is suggested
by ARPES measurements \cite{sato}, and the absence of impurity effect 
on $T_{\rm c}$ strongly indicates the realization of the 
$s$-wave superconducting state without sign change of the gap function.
In the present paper, we have studied the electron-phonon 
mechanism of superconductivity by considering two relevant optical 
phonon modes (breathing and shear phonons), and found that strong
pairing interaction is caused by the interband hopping
of Cooper pairs induced by shear phonons.
This mechanism is important
even if the top of $e_g'$ electron band is close to but below the Fermi level
as suggested experimentally.
Therefore, this valence-band SK mechanism is the origin of 
$s$-wave superconductivity in Na$_x$CoO$_2$, 
overcoming the strong Coulomb interaction $U\sim5$ eV.

In this paper, we have derived the EPI for breathing and shear phonons
by considering both the Coulomb potential and the transfer integrals.
The estimated EPI for shear phonon is prominently increased 
by water intercalation, resulting from the increase of 
trigonal distortion of CoO$_2$ layer.
According to the point charge model,
the top of the $e_g'$ bands $\Delta(<0)$ is expected to 
approach the Fermi level due to water intercalation.
Both effects induced by the water intercalation
will raise $T_{\rm c}$ due to the SK mechanism.

Based on the obtained model Hamiltonian, we determine $T_{\rm c}$ 
by solving the strong coupling Eliashberg equation.
The SK mechanism is seldom damaged by the Coulomb interaction
since the pair hopping $J$, which is the depairing force
for the SK mechanism, is much smaller than $U$.
For this reason, experimental $T_{\rm c}\sim 5$ K is realized
irrespective of the realistic Coulomb interaction $U\sim5$ eV.
We have also studied the oxygen isotope effect 
($^{16}{\rm O} \rightarrow ^{18}{\rm O}$) on $T_{\rm c}$.
In the absence of Coulomb interaction, 
$T_{\rm c}$ decreases by the isotope substitution
in proportion to $\omega_D\propto M_O^{-1/2}$.
However, we found that the isotope effect on $T_{\rm c}$ becomes 
very small for $U=4\sim6$ eV, since the renormalized
Coulomb interaction (Anderson-Morel potential) is reduced
with the decrease of $\omega_D$.

\acknowledgements
We are grateful to M. Sato, Y. Kobayashi, M. Yokoi and T. Moyoshi
for fruitful discussions on experimental results, 
including their unpublished data.
We are also grateful to G.-q. Zheng, K. Ishida, Y. Ihara, T. Shimojima,
H. Sakurai and  T. Sato for enlightening discussions on experiments.
Finally, we thank D.S. Hirashima, M. Ogata, K. Kuroki, Y. Tanaka,
Y. Yanase and M. Mochizuki for valuable comments and discussions
on theoretical issues.
This work was supported by the Grant-in-Aid for
Scientific Research from the Ministry of Education, 
Science, Sports and Culture of Japan.
Numerical calculations were performed at the supercomputer center, ISSP.

\appendix
\section{Analytical expression for the transition temperature in $\mbox{Na$_x$CoO$_2\cdot y$H$_2$O}$}
\begin{figure}[htbp]
\begin{center}
\includegraphics[scale=0.45]{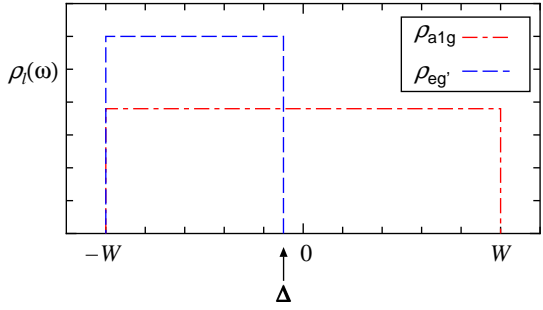}
\caption{(Color online) The simplified DOS for the $a_{1g}$- and $e_g'$-bands.}\label{DOS2}
\end{center}
\end{figure}

In this appendix, we derive the analytical expression for $s$-wave $T_c$ in Na$_x$CoO$_2\cdot y$H$_2$O.
For this purpose, we simplify the model further.
Here, we approximate the DOS for the $a_{1g}$- and $e_g'$-bands as 
$\rho_{a1g}(\omega)= \rho_{a1g}\theta(W-|\omega|)$ and
$\rho_{eg'}(\omega)= \rho_{eg'}\theta(\Delta-\omega)\theta(W+\omega)$,
which are shown in Fig. \ref{DOS2}.
Hereafter, we promise that $\Delta<0$.
We also approximate the phonon Green function by the step function
$D(\i\omega_m)=\frac{2}{\omega_{\rm D}}\theta(\omega_{\rm D}-|\omega_m|)$,
and assume that the Debye frequencies of breathing and shear phonon are the same for simplicity ($\omega_{\rm D}\equiv\omega_{\rm BR}=\omega_{\rm SH}$).
Then, the gap equation at $T=T_{\rm c}$ given in Eq. (\ref{eq:eliash}) becomes
\begin{eqnarray}
\hat\phi(\i\varepsilon_n)&=&T_{\rm c}\sum_{|\omega_m|\le\omega_{\rm D}}
\left( 
\begin{array}{cc}
a_1^2 &2b_1^2 \\
b_1^2 &a_2^2+2b_2^2 \\
\end{array} 
\right)\frac{2}{\omega_{\rm D}}
\hat F(\i\varepsilon_n-\i\omega_m)\nonumber\\
&&{}-T_{\rm c}\sum_{\omega_m}
\left( 
\begin{array}{cc}
U &2J \\
J &U+J \\
\end{array}
\right)
\hat F(\i\varepsilon_n-\i\omega_m),\label{eq:gap}
\end{eqnarray}
Here, we assume that $\hat\phi(\i\varepsilon_n)$ is given by the step function
\begin{eqnarray}
\hat\phi(\i\varepsilon_n)&=&
\left\{
\begin{array}{cc}
\hat\phi^{\rm I} &(|\varepsilon_n|<\omega_{\rm D}) \\
\hat\phi^{\rm II} &(|\varepsilon_n|>\omega_{\rm D})\\
\end{array}
\right..
\end{eqnarray}
First, we consider the case of $|\varepsilon_n|\ll\omega_{\rm D}$.
Then, the gap equation for $\hat\phi^{\rm I}$ is obtained from Eq. (\ref{eq:gap}).
\begin{widetext}
\begin{eqnarray}
\hat\phi^{\rm I}&=&
\left[
\left( 
\begin{array}{cc}
a_1^2 &2b_1^2 \\
b_1^2 &a_2^2+2b_2^2 \\
\end{array} 
\right)\frac{2}{\omega_{\rm D}}
-\left( 
\begin{array}{cc}
U &2J \\
J &U+J \\
\end{array}
\right)
\right]
\left( 
\begin{array}{cc}
\sum^{\rm I}|G_{a1g}({\bf k},{\rm i}\varepsilon_\ell)|^2\phi^{\rm I}_{a1g} \\
\sum^{\rm I}|G_{eg'}({\bf k},{\rm i}\varepsilon_\ell)|^2\phi^{\rm I}_{eg'} \\
\end{array}
\right)
\nonumber\\
&&{}-
\left( 
\begin{array}{cc}
U &2J \\
J &U+J \\
\end{array}
\right)
\left( 
\begin{array}{cc}
\sum^{\rm II}|G_{a1g}({\bf k},{\rm i}\varepsilon_\ell)|^2\phi^{\rm II}_{a1g} \\
\sum^{\rm II}|G_{eg'}({\bf k},{\rm i}\varepsilon_\ell)|^2\phi^{\rm II}_{eg'} \\
\end{array}
\right),\label{eq:gap1}
\end{eqnarray}
\end{widetext}
where $\sum^{\rm I}$ and $\sum^{\rm II}$
represent $\frac{T_{\rm c}}{N}\sum_{\bf k}\sum_{|\varepsilon_\ell|\le\omega_{\rm D}}$
and $\frac{T_{\rm c}}{N}\sum_{\bf k}\sum_{|\varepsilon_\ell|>\omega_{\rm D}}$, respectively.
Next, we consider the case of $|\varepsilon_n| \gg \omega_{\rm D}$,
where $F(i\epsilon_n-i\omega_m)\propto 
(\phi^{\rm II}_{a1g},\phi^{\rm II}_{eg'})$ for $|\omega_m|\le \omega_{\rm D}$.
Then, the gap equation for $\hat\phi^{\rm II}$ is given by
\begin{eqnarray}
\hat\phi^{\rm II}
&\simeq&\
-\left( 
\begin{array}{cc}
U &2J \\
J &U+J \\
\end{array}
\right)
\left( 
\begin{array}{cc}
\sum^{\rm II}|G_{a1g}({\bf k},{\rm i}\varepsilon_\ell)|^2\phi^{\rm II}_{a1g} \\
\sum^{\rm II}|G_{eg'}({\bf k},{\rm i}\varepsilon_\ell)|^2\phi^{\rm II}_{eg'} \\
\end{array}
\right)
\nonumber\\
&&{}-\left( 
\begin{array}{cc}
U &2J \\
J &U+J \\
\end{array}
\right)
\left( 
\begin{array}{cc}
\sum^{\rm I}|G_{a1g}({\bf k},{\rm i}\varepsilon_\ell)|^2\phi^{\rm I}_{a1g} \\
\sum^{\rm I}|G_{eg'}({\bf k},{\rm i}\varepsilon_\ell)|^2\phi^{\rm I}_{eg'} \\
\end{array}
\right). \nonumber \\ \label{eq:gap2}
\end{eqnarray}
Here, we ignored the phonon contribution
that is proportional to $\phi^{\rm II}$,
since it is much smaller than the first term in Eq. (\ref{eq:gap2}).

Using eq. (\ref{eq:gap2}), 
$\hat\phi^{\rm II}$ in Eq. (\ref{eq:gap1}) can be eliminated.
Then, the obtained gap equation for $\hat\phi^{\rm I}$ is given by
\begin{eqnarray}
\hat\phi^{\rm I}&=&\hat V^*_{\rm eff}\left( 
\begin{array}{cc}
\sum^{\rm I}|G_{a1g}({\bf k},{\rm i}\varepsilon_\ell)|^2\phi^{\rm I}_{a1g} \\
\sum^{\rm I}|G_{eg'}({\bf k},{\rm i}\varepsilon_\ell)|^2\phi^{\rm I}_{eg'} \\
\end{array}
\right),\label{eq:eliash2}\\
V^*_{\rm eff}&=&\left( 
\begin{array}{cc}
a_1^2 &2b_1^2 \\
b_1^2 &a_2^2+2b_2^2 \\
\end{array} 
\right)\frac{2}{\omega_{\rm D}}
 \nonumber \\
& &-\left( 
\begin{array}{cc}
U^*_{a1g} &2J^*_{a1g,eg'} \\
J^*_{a1g,eg'} &U^*_{eg'}+J^*_{eg',eg'} \\
\end{array}
\right), 
\end{eqnarray}
where $\hat V^*_{\rm eff}$ is the renormalized effective interaction.
$U^*_{a1g}$ ($U^*_{eg'}$) is the renormalized Coulomb repulsion for the $a_{1g}$ ($e_g'$) orbital,
and $J^*_{a1g,eg'}$ ($J^*_{eg',eg'}$) is the renormalized pair hopping
between ($a_{1g}$, $e_g'$) orbitals (($e_g'$, $e_g'$) orbitals).
These renormalized Coulomb interactions satisfy the following simultaneous equations \cite{yada3}.
\begin{eqnarray}
U^*_{a1g}&=&U-UAU^*_{a1g}-2JBJ^*_{a1g,eg'},\label{eq1}\\
U^*_{eg'}&=&U-UBU^*_{eg'}-JAJ^*_{a1g,eg'}
\nonumber \\
& &-JBJ^*_{eg',eg'},
\end{eqnarray}
\begin{eqnarray}
J^*_{a1g,eg'}&=&J-JBU^*_{eg'}-UAJ^*_{a1g,eg'}
\nonumber \\
& &-JBJ^*_{eg',eg'},\\
J^*_{eg',eg'}&=&J-JBU^*_{eg'}-JAJ^*_{a1g,eg'}
\nonumber \\
& &-UBJ^*_{eg',eg'},\label{eq4}
\end{eqnarray}
where $A=\sum^{\rm II}|G_{a1g}({\bf k},{\rm i}\varepsilon_\ell)|^2$ and $B=\sum^{\rm II}|G_{eg'}({\bf k},{\rm i}\varepsilon_\ell)|^2$.
If we assume $|\Delta|\ll\omega_{\rm D}$, then
$A=\rho_{a1g}\ln(W/\omega_{\rm D})$ and $B=\rho_{eg'}\int_{\omega_{\rm D}}^W\frac{1}{\varepsilon}(\frac{1}{2}+\frac{1}{\pi}{\rm Arctan}(\frac{\Delta}{\varepsilon}))d\varepsilon\approx\frac{1}{2}\rho_{eg'}\ln(W/\omega_{\rm D})$.

We now calculate $T_{\rm c}$ from Eq. (\ref{eq:eliash2}).
In the case of $|\Delta|\ll\omega_{\rm D}\ll W$,
$\sum^{\rm I}|G_{\ell}({\bf k},i\varepsilon_n)|^2$ is approximated as follows.
\begin{eqnarray}
& &
\frac{T_{\rm c}}{N}\sum_{\bf k}\sum_{|\varepsilon_n|<\omega_{\rm D}}|G_{a1g}({\bf k},i\varepsilon_n)|^2
 \nonumber \\
&\simeq&T_{\rm c}\sum_{|\varepsilon_n|<\omega_{\rm D}}\int_{-\infty}^{\infty}\frac{z_{a_{1g}}\rho_{a_{1g}}}{\varepsilon_n^2+E^2}dE\nonumber\\
&=&z_{a_{1g}}\rho_{a_{1g}}T_{\rm c}\sum_{|\varepsilon_n|<\omega_{\rm D}}\frac{\pi}{|\varepsilon_n|}\nonumber\\
&\simeq&z_{a_{1g}}\rho_{a_{1g}}\ln\left(\frac{1.13\omega_{\rm D}}{T_{\rm c}}\right).
\end{eqnarray}
\begin{eqnarray}
& &
\frac{T_{\rm c}}{N}\sum_{\bf k}\sum_{|\varepsilon_n|<\omega_{\rm D}}|G_{e_g'}({\bf k},i\varepsilon_n)|^2
\nonumber \\
&\simeq& T_{\rm c}\sum_{|\varepsilon_n|<\omega_{\rm D}}\int_{-\infty}^{\Delta}\frac{z_{e_g'}\rho_{e_g'}}{\varepsilon_n^2+E^2}dE\nonumber\\
&=&z_{e_g'}\rho_{e_g'}T_{\rm c}\sum_{|\varepsilon_n|<\omega_{\rm D}}\frac{\pi}{|\varepsilon_n|}
\left(\frac{1}{2}+\frac{1}{\pi}{\rm Arctan}\left(\frac{\Delta}{|\varepsilon_n|}\right)\right)\nonumber\\
&\simeq&\rho_{e_g'}z_{e_g'}\left(\frac{|\Delta|}{\pi\omega_{\rm D}}
+\frac{1}{2}\ln\left(\frac{\omega_{\rm D}}{|\Delta|}\right)\right).
\end{eqnarray}
As a result, according to Eq. (\ref{eq:eliash2}),
the equation for $T_{\rm c}$ is obtained as
\begin{widetext}
\begin{eqnarray}
\frac{\left(\lambda^*_1\ln\frac{1.13\omega_{\rm D}}{T_{\rm c}}+\frac{\lambda^*_4}{\lambda_\Delta}\right)+
\sqrt{\left(\lambda^*_1\ln\frac{1.13\omega_{\rm D}}{T_{\rm c}}-\frac{\lambda^*_4}{\lambda_\Delta}\right)^2
+\frac{4\lambda^*_2\lambda^*_3}{\lambda_\Delta}\ln\frac{1.13\omega_{\rm D}}{T_{\rm c}}}}{2}=1, \label{Tc-ap}
\end{eqnarray}
where
$\lambda^*_1=\left(\frac{2a_1^2}{\omega_{\rm D}}-U_{a_{1g}}^*\right)z_{a_{1g}}\rho_{a_{1g}}$,
$\lambda^*_2=\left(\frac{4b_1^2}{\omega_{\rm D}}-2J_{a_{1g},e_g'}^*\right)z_{e_g'}\rho_{e_g'}$,
$\lambda^*_3=\left(\frac{2b_1^2}{\omega_{\rm D}}-J_{a_{1g},e_g'}^*\right)z_{a_{1g}}\rho_{a_{1g}}$,
$\lambda^*_4=\left(\frac{2(a_2^2+2b_2^2)}{\omega_{\rm D}}-U_{e_g'}^*-J_{e_g',e_g'}^*\right)z_{e_g'}\rho_{e_g'}$
and
$\lambda_\Delta=\left(\frac{|\Delta|}{\pi\omega_{\rm D}}+\frac{1}{2}\ln\left(\frac{\omega_{\rm D}}{|\Delta|}\right)\right)^{-1}$.
\end{widetext}
By solving eq. (\ref{Tc-ap}), $T_c$ is given by
\begin{eqnarray}
T_{\rm c}&=&1.13\omega_{\rm D}\exp\left(-\frac{1}{\lambda_{\rm eff}}\right), \label{eq:Tc} \\
\lambda_{\rm eff}&=&\lambda^*_1+\frac{\lambda^*_2\lambda^*_3}{\lambda_\Delta-\lambda^*_4}.\label{eq:lambda}
\end{eqnarray}
In Eq. (\ref{eq:lambda}), the first term originates from breathing phonon,
and the second term originates from valence band SK effect.
$\lambda_\Delta$ monotonically decreases with increasing $\Delta$.
Therefore, $T_c$ increase as the top of the $e_g'$ band approaches the Fermi level.



\begin{thebibliography}{99}

\bibitem{takada}
K. Takada, H. Sakurai, E. Takayama-Muromachi, F. Izumi, R. A. Dilanian and T. Sasaki:
Nature {\bf 422} (2003) 53.

\bibitem{foo}
M. L. Foo, Y. Wang, S. Watauchi, H. W. Zandbergen, T. He, R. J. Cava and N. P. Ong:
Phys. Rev. Lett. {\bf 92} (2004) 247001.

\bibitem{yokoi}
M. Yokoi, T. Moyoshi, Y. Kobayashi, M. Soda, Y. Yasui, M. Sato and K. Kakurai:
J. Phys. Soc. Jpn. 74 (2005) 3046.

\bibitem{hiroi}
D. Yoshizumi, Y. Muraoka, Y. Okamoto, Y. Kiuchi, J. Yamaura, M. Mochizuki, M. Ogata and Z. Hiroi:
J. Phys. Soc. Jpn. 76 (2007) 063705.

\bibitem{pes}
T. Shimojima, T. Yokoya, T. Kiss, A. Chainani, S. Shin, T. Togashi, S. Watanabe, C. Zhang, C. T. Chen,
K. Takada, T. Sasaki, H. Sakurai and E. Takayama-Muromachi: Phys. Rev. B 71 (2005) 020505(R).

\bibitem{wu}
D. Wu, J. L. Luo and N. L. Wang:
Phys. Rev. B {\bf 73} (2006) 014523.


\bibitem{kobayashi}
Y. Kobayashi, H. Watanabe, M. Yokoi, T. Moyoshi, Y. Mori and M. Sato:
J. Phys. Soc. Jpn. {\bf 74} (2005) 1800.

\bibitem{kobayashi_nmr}
Y. Kobayashi, T. Moyoshi, H. Watanabe, M. Yokoi and M. Sato:
J. Phys. Soc. Jpn. {\bf 75} (2006) 074717.

\bibitem{zheng_nmr}
G.-q. Zheng, K. Matano, D. P. Chen and C. T. Lin:
Phys. Rev. B {\bf 73} (2006) 180503(R).

\bibitem{fujimoto}
T. Fujimoto, G.-q. Zheng, Y. Kitaoka, R. L. Meng, J. Cmaidalka and C.W. Chu:
Phys. Rev. Lett. {\bf 92} (2004) 047004.

\bibitem{ishida}
K. Ishida, Y. Ihara, Y. Maeno, C. Michioka, M. Kato, K. Yoshimura, K. Takada,
T. Sasaki, H. Sakurai and E. Takayama-Muromachi:
J. Phys. Soc. Jpn. {\bf 72} (2003) 3041.

\bibitem{zheng_nqr}
G.-q. Zheng, K. Matano, R. L. Meng, J. Cmaidalka and C. W. Chu:
J. Phys.: Condens. Matter {\bf 18} (2006) L63.

\bibitem{yokoi_imp}
M. Yokoi, H. Watanabe, Y. Mori, T. Moyoshi, Y. Kobayashi and M. Sato:
J. Phys. Soc. Jpn. {\bf 73} (2004) 1297.

\bibitem{singh}
D. J. Singh:
Phys. Rev. B {\bf 61} (2000) 13397.

\bibitem{yang}
H.-B. Yang, Z.-H. Pan, A. K. P. Sekharan, T. Sato, S. Souma, T. Takahashi, R. Jin, B. C. Sales,
D. Mandrus,A. V. Fedorov, Z. Wang and H. Ding: Phys. Rev. Lett. 95 (2005) 146401
\bibitem{sato}
T. Arakane, T. Sato, T. Takahashi, H. Ding, T. Fujii and Atsushi Asamitsu:
J. Phys. Soc. Jpn. {\bf 76} (2007) 054704.

\bibitem{shimojima}
T. Shimojima, K. Ishizaka, S. Tsuda, T. Kiss, T. Yokoya, A. Chainani, S. Shin,
P. Badica,K. Yamada and K. Togano: Phys. Rev. Lett. 97 (2006) 267003.

\bibitem{hasan}
D. Qian, L. Wray, D. Hsieh, L. Viciu, R. J. Cava, J. L. Luo, D. Wu, N. L. Wang, and M. Z. Hasan:
Phys. Rev. Lett. {\bf 97} (2006) 186405.

\bibitem{yada1}
K. Yada and H. Kontani:
J. Phys. Soc. Jpn. {\bf 74} (2005) 2161.

\bibitem{PALee}O. I. Motrunich and P. A. Lee:Phys. Rev. B {\bf 70} (2004) 024514. 

\bibitem{Baskaran}G. Baskaran:Phys. Rev. Lett. {\bf 91} (2003) 097003. 

\bibitem{Ogata-review}M. Ogata:J. Phys.: Condens. Matter {\bf 19} (2007) 145282.
\bibitem{nishikawa}
Y. Nisikawa, H. Ikeda and K. Yamada:
J. Phys. Soc. Jpn. {\bf 73} (2004) 1127.

\bibitem{yanase}
Y. Yanase, M. Mochizuki and M. Ogata:
J. Phys. Soc. Jpn. {\bf 74} (2005) 430.

\bibitem{mochizuki}
M. Mochizuki, Y. Yanase and M. Ogata:
Phys. Rev. Lett {\bf 94} (2005) 147005.

\bibitem{kuroki}
K. Kuroki, S. Onari, Y. Tanaka, R. Arita and T. Nojima:
Phys. Rev. B {\bf 73} (2006) 184503.

\bibitem{moyoshi}
T. Moyoshi, Y. Yasui, M. Soda, Y. Kobayashi,M. Sato and K. Kakurai:J. Phys. Soc. Jpn. {\bf 75} (2006) 074705

\bibitem{yada2}
K. Yada and H. Kontani:
J. Phys. Soc. Jpn. {\bf 75} (2006) 033705.

\bibitem{suhl}
H. Suhl, B. T. Matthias, and L. R. Walker: Phys. Rev. Lett. {\bf 3} (1959) 552;
J. Kondo: Prog. Theor. Phys. {\bf 29} (1963) 1. 

\bibitem{koshibae}
W. Koshibae and S. Maekawa:
Phys. Rev. Lett {\bf 91} (2003) 257003.

\bibitem{ionic}
W. B. Wu, D. J. huang, J. Okamoto, A. Tanaka, H.-J. Lin, F. C. Chou. A. Fujimori and C. T. Chen:
Phys. Rev. Lett {\bf 94} (2005) 146402.

\bibitem{kukoki-s}K. Kuroki, S. Onari, Y. Tanaka, R. Arita and T. Nojima:Phys. Rev. B {\bf 73} (2006) 184503. 

\bibitem{mochizuki-s}M. Mochizuki and M. Ogata:J. Phys. Soc. Jpn {\bf 76} (2007) 013704.

\bibitem{landron}
S. Landron and M.-B. Lepetit:
cond-mat/0605454.

\bibitem{oxonium}
H. Sakurai, N. Tsujii, O. Suzuki, H. Kitazawa, G. Kido, K. Takada, T. Sasaki and E. Takayama-Muromachi:
Phys. Rev. B {\bf 74} (2006) 092502.

\bibitem{li}
Z. Li, J. Yang, J. G. Hou and Q. Zhu:
Phys. Rev. B {\bf 70} (2004) 144518.

\bibitem{Harrison}
 W. A. Harrison:
 {\it Elementary Electronic Structure}
 (World Scientific, 1999, Singapore).

\bibitem{slater}
J. C. Slater and G. F. Koster:
Phys. Rev. {\bf 94} (1954) 1498

\bibitem{lynn}
J. W. Lynn, Q. Huang, C. M. Brown, V. L. Miller, M. L. Foo, R. E. Schaak, C. Y. Jones,
E. A. Mackey and R. J. Cava:
Phys. Rev. B {\bf 68} (2003) 214516

\bibitem{radius}
{\it e.g.,} R.D. Shannon and C.T.  Prewitt,
Acta Crystallogr. B {\bf 25}, 925 (1969).

\bibitem{frequency}
P. Lemmens, K. Y. Choi, V. Gnezdilov, E. Ya. Sherman, D. P. Chen, C. T. Lin, F. C. Chou and B. Keimer:
Phys. Rev. Lett. {\bf 96} (2006) 167204.

\bibitem{gutz}S. Zhou, M. Gao, H. Ding, P. A. Lee and Z. Wang:Phys. Rev. Lett. {\bf 94} (2005) 206401.


\bibitem{dmfa1}H. Ishida, M. D. Johannes and A. Liebsch:Phys. Rev. Lett. {\bf 94} (2005) 196401.

\bibitem{dmfa2}C. A. Marianetti, K. Haule and O. Parcollet:Phys. Rev. Lett. {\bf 99} (2007) 246404.

\bibitem{morel}
P. Morel and P. W. Anderson: Phys. Rev. 125 (1962) 1263.

\bibitem{yada3}
K. Yada and H. Kontani:
Journal of Magnetism and Magnetic Materials {\bf 310} (2007) 684.

\bibitem{yokoi_isotope}
M. Yokoi, Y. Kobayashi and M. Sato: unpublished.

\bibitem{Cv1}
H. D. Yang, J.-Y. Lin, C. P. Sun, Y. C. Kang, C. L. Huang, K. Takada, T. Sasaki, H. Sakurai and E.
Takayama-Muromachi:
Phys. Rev. B {\bf 71} (2005) 020504(R).

\bibitem{Cv2}
N. Oeschler, R. A. Fisher, N. E. Phillips, J. E. Gordon, M.-L. Foo and R. J. Cava:
cond-mat/0503690.

\bibitem{Izawa}
K. Izawa, K. Kamata, Y. Nakajima, Y. Matsuda, T. Watanabe, N. Nohara, H. Takagi, P. Thalmeier, and K. Maki, Phys. Rev. Lett. {\bf 89}, 137006 (2002).

\bibitem{watanabe}
 T. Watanabe, M. Nohara, T. Hanaguri and H. Takagi:
Phys. Rev. Lett {\bf 92} (2004) 147002.

\bibitem{Nohara}
 M. Nohara, M. Isshiki, F. Sakai and H. Takagi:
 J. Phys. Soc. Jpn {\bf 68} (1999) 1078.

\bibitem{kontani}
H. Kontani: Phys. Rev. B 70 (2004) 054507.

\bibitem{ishida2}
Y. Ihara, H. Takeya, K. Ishida, H. Ikeda, C. Michioka, K. Yoshimura,
K. Takada, T. Sasaki, H. Sakurai and E. Takaayama-Muromachi:
J. Phys. Soc. Jpn. 75 (2006) 124714.

\bibitem{moyoshi-private} T. Moyoshi and M. Sato, private communication.

\bibitem{kontani2}
present authors, unpublished.

\end{thebibliography}
\end{document}